\documentclass[showpacs, aps,prd,twocolumn,superscriptaddress,preprintnumbers]{revtex4-1}
\usepackage{units}
\usepackage[backgroundcolor=green!40, disable]{todonotes}
\usepackage{amsmath,amssymb}
\usepackage{lineno}
\usepackage{subscript}
\usepackage{xspace}
\usepackage{graphicx}
\usepackage{fancyhdr}
\usepackage{subfigure}
\usepackage{times}
\usepackage{gensymb} 
\usepackage{afterpage}
\newcommand{\Xmax}{\ensuremath X_\mathrm{max}}

\newcommand{\rt}{\ensuremath t_{1/2}}
\newcommand{\meanrt}{\ensuremath \langle t_{1/2}/r\rangle}
\newcommand{\secMax}{\ensuremath(\sec \theta)_\mathrm{max}}
\newcommand{\secMaxp}{\ensuremath(\sec \theta)_\mathrm{max;p}}
\newcommand{\secMaxFe}{\ensuremath(\sec \theta)_\mathrm{max;Fe}}
\newcommand{\MeanlnA}{\ensuremath \langle \ln A \rangle}
\newcommand{\lnA}{\ensuremath \ln A}

\begin{document}
\preprint{Published in Phys. Rev. D as DOI:http://dx.doi.org/10.1103/PhysRevD.93.072006}
\title{Azimuthal asymmetry in the risetime of the surface detector signals of the Pierre Auger Observatory}
%

\author{A.~Aab}
\affiliation{Universit\"at Siegen, Fachbereich 7 Physik -- Experimentelle Teilchenphysik, Germany}

\author{P.~Abreu}
\affiliation{Laborat\'orio de Instrumenta\c{c}\~ao e F\'\i{}sica Experimental de Part\'\i{}culas -- LIP and Instituto Superior T\'ecnico -- IST, Universidade de Lisboa -- UL, Portugal}

\author{M.~Aglietta}
\affiliation{Osservatorio Astrofisico di Torino (INAF), Torino, Italy}
\affiliation{INFN, Sezione di Torino, Italy}

\author{E.J.~Ahn}
\affiliation{Fermi National Accelerator Laboratory, USA}

\author{I.~Al Samarai}
\affiliation{Laboratoire de Physique Nucl\'eaire et de Hautes Energies (LPNHE), Universit\'es Paris 6 et Paris 7, CNRS-IN2P3, France}

\author{I.F.M.~Albuquerque}
\affiliation{Universidade de S\~ao Paulo, Inst.\ de F\'\i{}sica, S\~ao Paulo, Brazil}

\author{I.~Allekotte}
\affiliation{Centro At\'omico Bariloche and Instituto Balseiro (CNEA-UNCuyo-CONICET), Argentina}

\author{P.~Allison}
\affiliation{Ohio State University, USA}

\author{A.~Almela}
\affiliation{Instituto de Tecnolog\'\i{}as en Detecci\'on y Astropart\'\i{}culas (CNEA, CONICET, UNSAM), Centro At\'omico Constituyentes, Comisi\'on Nacional de Energ\'\i{}a At\'omica, Argentina}
\affiliation{Universidad Tecnol\'ogica Nacional -- Facultad Regional Buenos Aires, Argentina}

\author{J.~Alvarez Castillo}
\affiliation{Universidad Nacional Aut\'onoma de M\'exico, M\'exico}

% ...

\collaboration{The Pierre Auger Collaboration}
\email{auger\_spokespersons@fnal.gov}
\homepage{http://www.auger.org}
\noaffiliation

%
\begin{abstract}
The azimuthal asymmetry in the risetime of signals in Auger surface detector stations is a source of information on shower development. 
The azimuthal asymmetry is due to a combination of the longitudinal evolution of the shower and geometrical effects related to the angles of incidence of 
the particles into the detectors. The magnitude of the effect depends upon the zenith angle and state of development of the shower and thus provides a novel observable, $\secMax$, sensitive to the mass composition of cosmic rays above $3 \times 10^{18}$ eV. 
By comparing measurements with predictions from shower simulations, we find for both of our adopted models of hadronic physics (QGSJETII-04 and EPOS-LHC) an indication that the mean cosmic-ray mass increases slowly with energy, as has been inferred from other studies. However, the mass estimates are dependent on the shower model and on the range of distance from the shower core selected. Thus the method has uncovered further deficiencies in our understanding of shower modelling that must be resolved before the mass composition can be inferred from $\secMax$.
\end{abstract}
\maketitle
%
%
%

\section{\label{sec:Introduction}Introduction}

A detailed understanding of the properties and origin of cosmic rays with energies greater than 1 Joule ($6.3 \times 10^{18}$ eV) remains incomplete over 50 years since their discovery \cite{PhysRevLett.6.485}. An explanation for the origin of these particles 
must account for the observations of their energy spectrum, arrival direction distributions and mass composition. Determination of the mass composition from measurements of extensive air showers is by far the greatest challenge as it is necessary to make 
assumptions about the hadronic physics in regions of phase space not covered by measurements at accelerators: for example, the center-of-mass energy that will ultimately be reached at the LHC corresponds to that reached in a collision of a proton of only $10^{17}$ eV 
with a stationary nucleon. It follows that in the region of interest here, from $10^{18}$ to $10^{20}$ eV, there is a serious lack of knowledge of how key parameters such as the cross-section, the multiplicity and the inelasticity in collisions of protons and nuclei 
on nuclei, and of charged pions on nuclei, depend on energy. Furthermore, at the LHC, studies are restricted to relatively small rapidities whereas at air-shower energies the behavior at large Feynman $x$ is of great significance.

Lack of knowledge of the hadronic processes is a less serious issue when deriving the energy spectrum following the successful demonstration of calorimetric estimates of the energies of cosmic rays using the fluorescence technique \cite{Abbasi:2007sv,Abraham:2008ru}. 
In determining the energy account must be taken of the ``invisible energy'' carried by neutrinos and by muons taken into the earth beyond the reach of the fluorescence detectors and the unknowns of mass composition and hadronic physics are important at about the 
10\% level. Methods are also being developed to estimate the invisible energy on an event-by-event basis \cite{Tueros:ICRC13}. In \cite{Abbasi:2007sv,Abraham:2008ru} convincing evidence for a suppression of the spectrum flux above $\sim 4 \times 10^{19}$ eV was 
reported. However, to interpret the shape of the spectrum in detail requires knowledge of the mass composition at the highest energies.

The fluorescence technique can be used to get information that relates to the mass composition.  It has been used to measure the average depth and spread of the distribution of points at which the number of particles in the shower maximizes, $\Xmax$, as a function of energy. 
Measurements of $\Xmax$ based on observations of nearly 20000 events above $6.3 \times 10^{17}$ eV have recently been reported \cite{Aab:2014kda}. However only 37 of these events have energies above $3.2 \times 10^{19}$ eV, a number constrained by the on-time of 
fluorescence detectors which is $< 13\%$. Detailed analyses of the distributions of $\Xmax$ in narrow energy bins, made using three models of the hadronic interaction, Sibyll 2.1 \cite{Fletcher:1994bd}, QGSJETII-04 \cite{Ostapchenko:2010vb} and 
EPOS-LHC \cite{Pierog:2013ria}, lead to the conclusion that helium and nitrogen are the most abundant elements above $3.2 \times 10^{19}$ eV \cite{Aab:2014aea}.

The lack of compositional information at the highest energies is also a severe problem for the interpretation of the distributions of arrival directions. For example the high degree of isotropy observed in numerous tests of the small-scale angular distributions 
of ultra-high energy cosmic rays (UHECR) is remarkable
, challenging earlier expectations that assumed only a few cosmic-ray sources producing light primaries at the highest energies. In fact the largest departures from isotropy are observed for cosmic rays above 
$5.8 \times 10^{19}$ eV in sky-windows of about 20$\degree$ \cite{PierreAuger:2014yba}, while at energies above 8 EeV, there are indications of a dipole in the distribution of arrival directions \cite{ThePierreAuger:2014nja}.

One way to increase the sample, and so test the interaction models, is to develop techniques using the water-Cherenkov detectors of the surface array of the Auger Observatory \cite{Aab:2015zoa}, which operate $\sim$ 100\% of the time. It has been shown that the 
depth of production of muons (MPD) \cite{MPD:2014dua} contains relevant information on mass composition up to energies beyond $6 \times 10^{19}$ eV. However the variation of mass with energy, deduced when the depth of maximum of muon production ($X_{\mu}^{\rm max}$) is compared to 
the predictions of the QGSJETII-04 and EPOS-LHC hadronic models, does not agree with what is found from the fluorescence detector (FD) measurements suggesting that the part of the hadronic development that relates to muon creation is modelled incorrectly.

In this paper a further exploration of the model-mass parameter space is described using an observable from the water-Cherenkov detectors that is related to the azimuthal asymmetry found in the risetime of the signals with respect to the direction of the 
incoming air shower. The asymmetry is due to a combination of the longitudinal development of the shower and of geometrical effects and thus has the potential to give alternative insights into the matching of data to mass and hadronic models using a technique 
with quite different systematic uncertainties to those of the MPD or FD approaches. As both the muonic and electromagnetic components contribute to the risetime, the technique explores the region between the dominantly electromagnetic study of $\Xmax$ and the MPD 
analysis which is muon-based.

The structure of the paper is as follows. In the following section the Auger Observatory is briefly outlined with emphasis on aspects that are important for this paper. In section~\ref{sec:Method} the concept of the asymmetry of the risetime is described while in 
section~\ref{sec:AnalysisTitle} details of the analysis of this asymmetry are presented. The results are shown in section~\ref{sec:results} and discussed in section~\ref{sec:Conclusions} where they are compared with those from the fluorescence detector and the MPD analyses.

\begin{figure}[t!]
     \centerline{\includegraphics[width=0.5\textwidth]{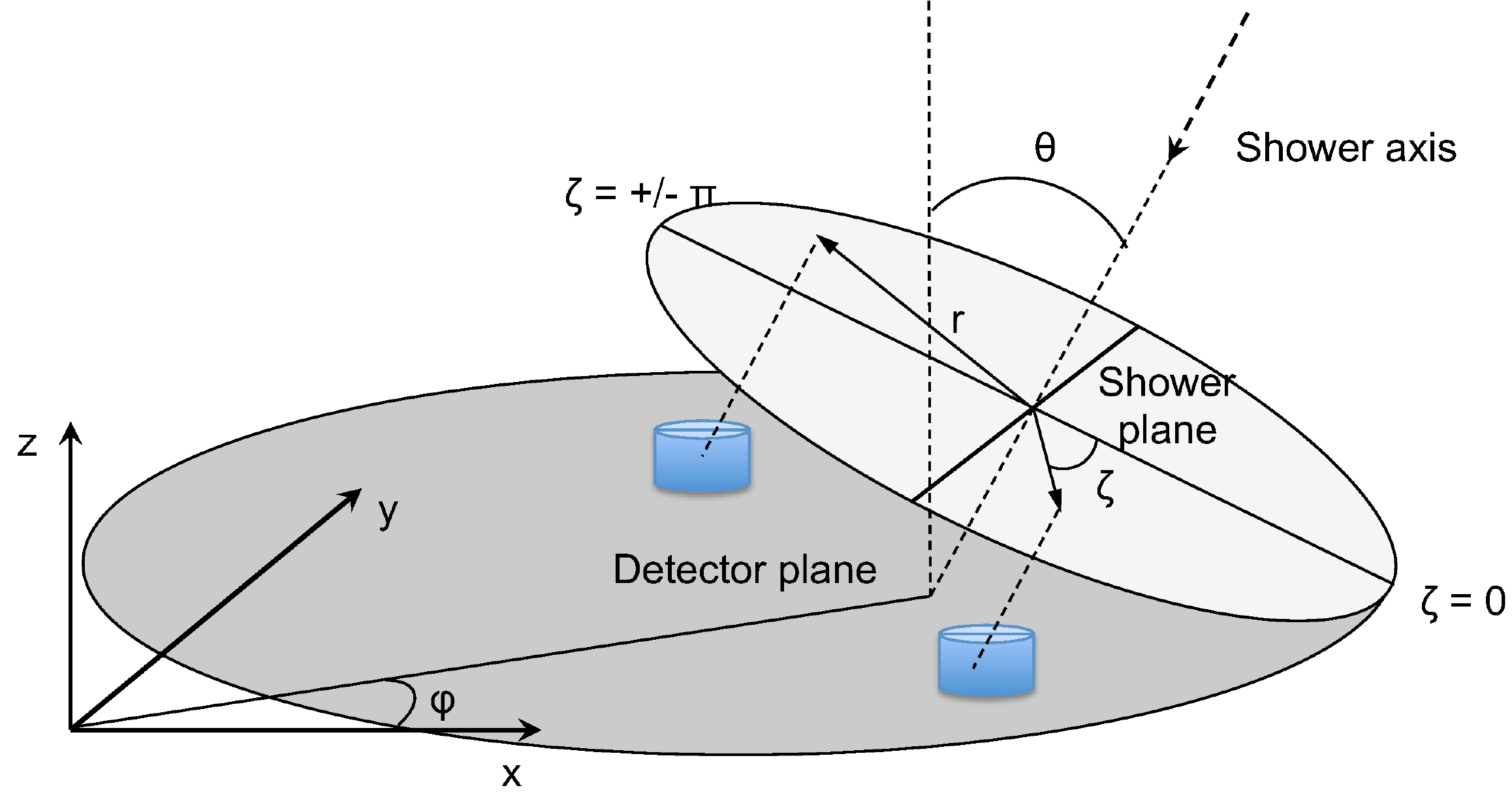}}
     \caption{Schematic view of the shower geometry. The incoming direction of the primary particle defines two regions, ``early'' ($\mid\zeta\mid \,< \pi/2$) and ``late'' region ($\mid\zeta\mid \,> \pi/2$). Note the different amount of atmosphere traversed by the 
particles reaching the detectors in each region.}
     \label{fig:CoordinateSystem}
 \end{figure}

\section{\label{sec:Observatory}The Observatory and event reconstruction}

The Pierre Auger Observatory is located in the Province of Mendoza, Argentina (35.1$\degree$- 35.5$\degree$S, 69.0$\degree$- 69.6$\degree$W, 1400 m a.s.l.). It is a hybrid system, a combination of a large surface-detector array (SD) and a fluorescence detector which records cosmic-ray events above $10^{17}$ eV. The work presented in the following is based on data from the SD. As data from the FD will also be referred to, we summarize here the main characteristics of the two detectors as well as the principles 
of the event reconstruction. More details on the detectors and on the event reconstruction can be found in \cite{Aab:2015zoa,Abraham:2004dt,Abraham:2010mj,Abraham:2010yv}. 

\begin{figure*}[t]
    \includegraphics[width=0.40\textwidth]{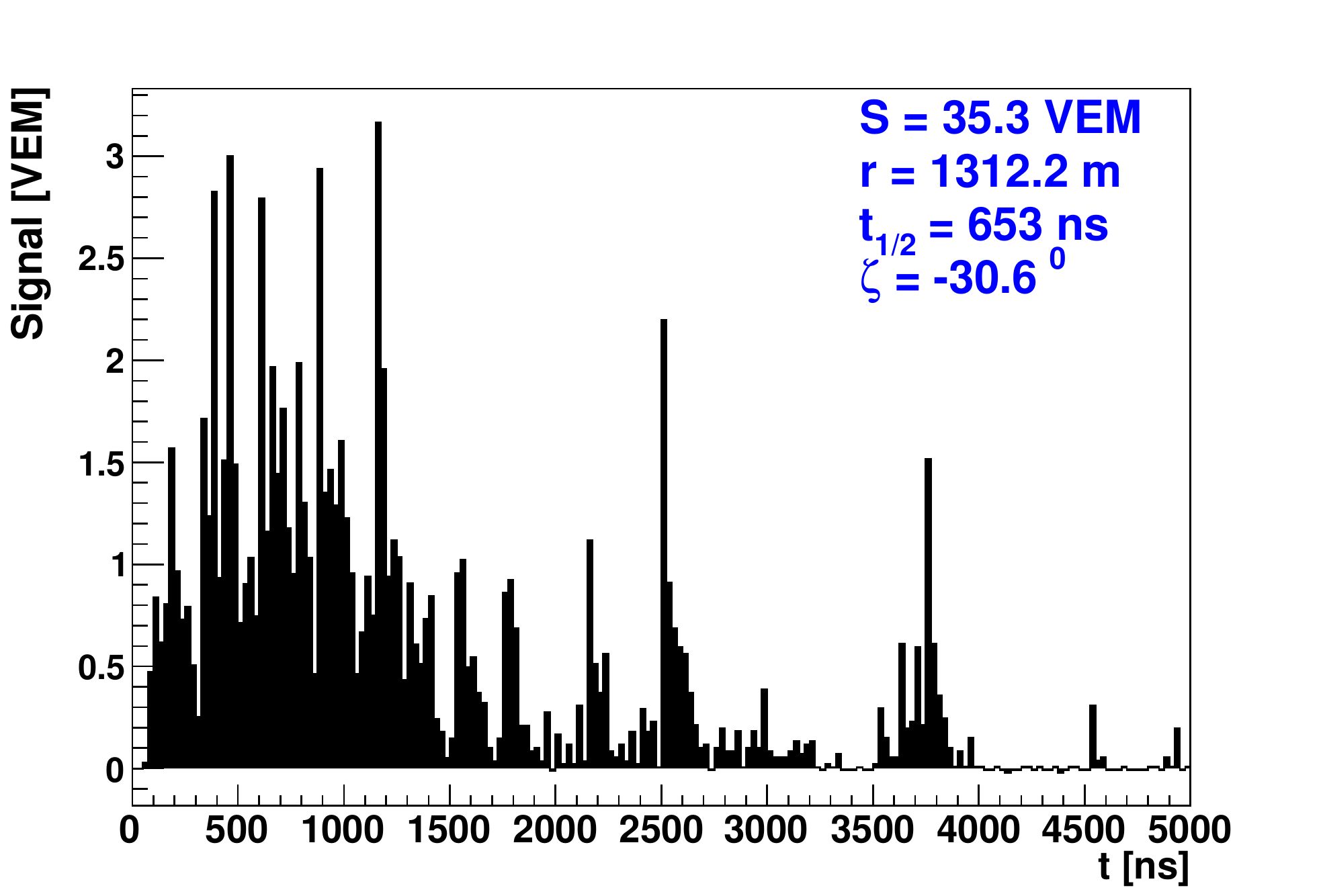}
    \hspace{0.05\textwidth}
    \includegraphics[width=0.40\textwidth]{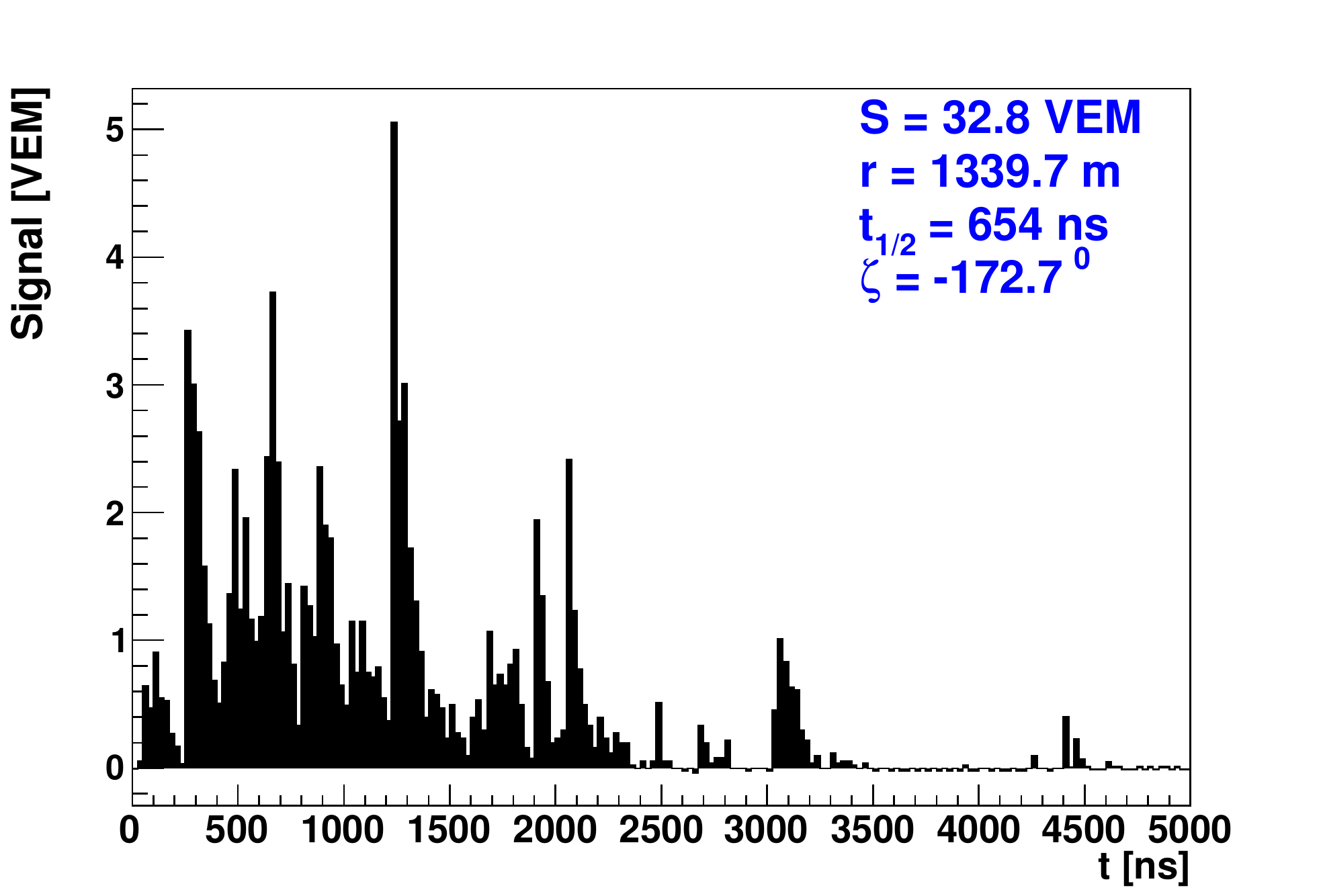}
    \includegraphics[width=0.40\textwidth]{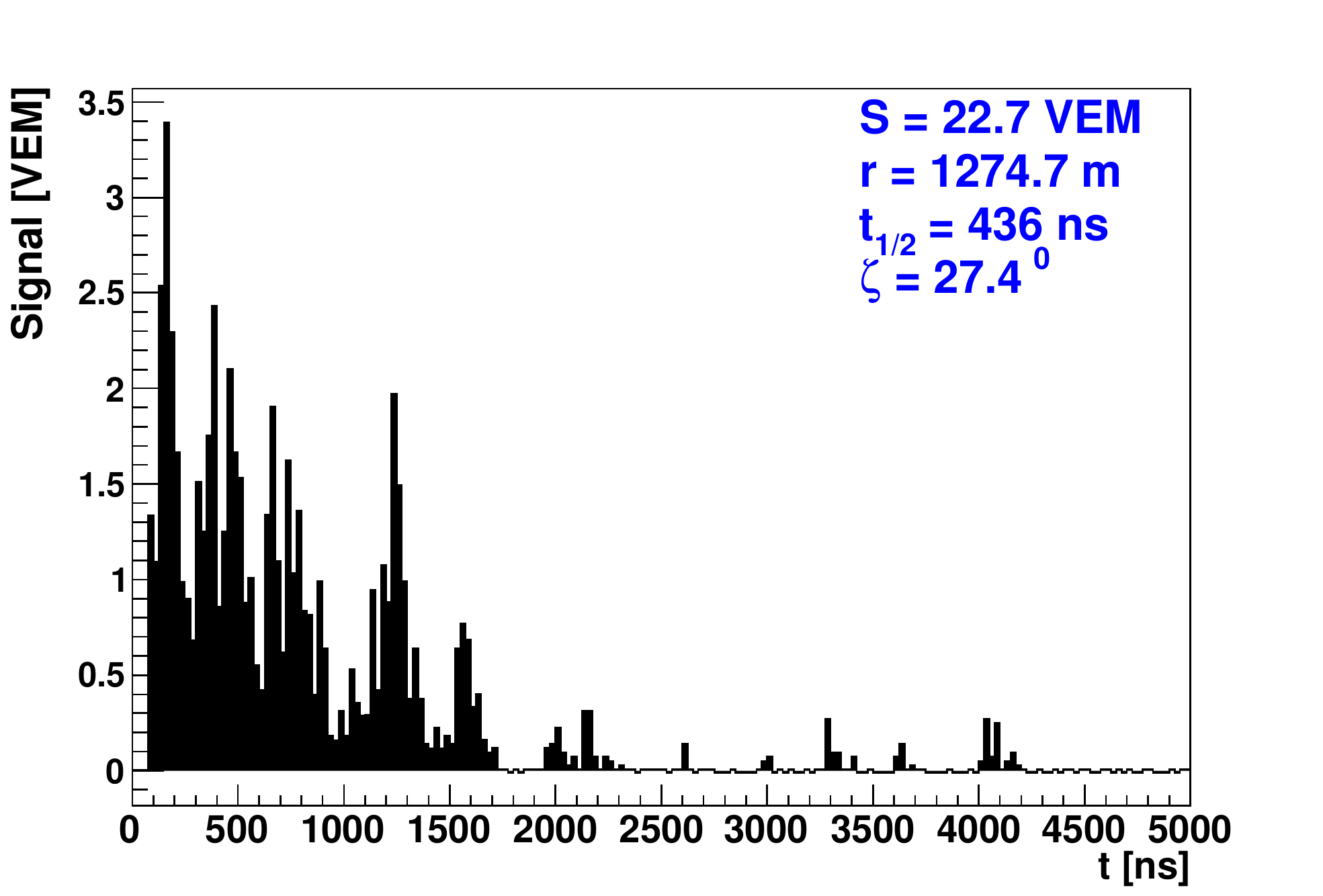}
    \hspace{0.05\textwidth}
    \includegraphics[width=0.40\textwidth]{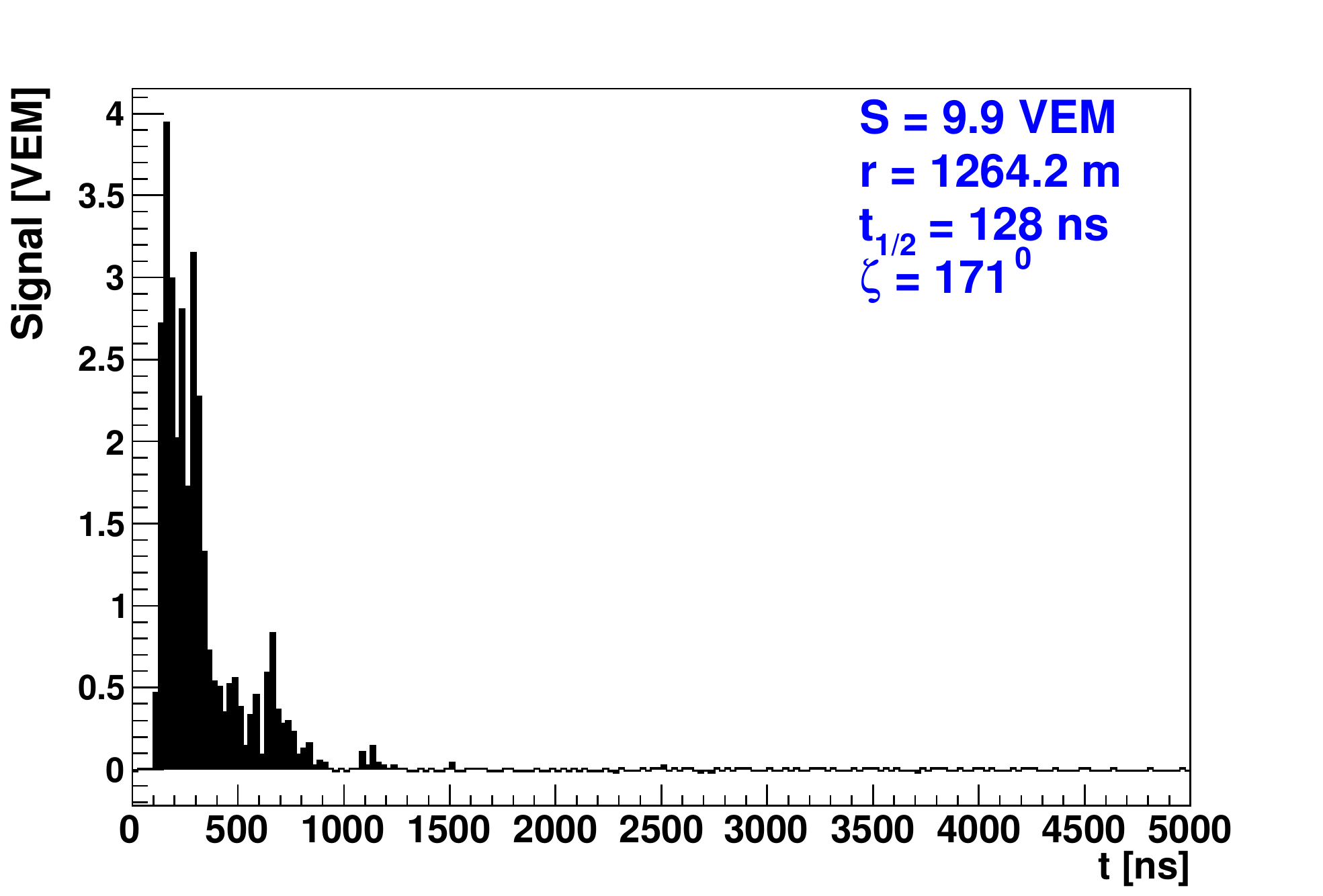}
\caption{Top: two stations in an event of 16.9 EeV and 15.7$\degree$ in zenith. Bottom: two stations in an event of 7.7 EeV and 52$\degree$ in zenith. Left panels correspond to early stations while right panels correspond to late stations.}
\label{fig:event_stations}
\end{figure*}

The surface detector array, covering an area of over 3000 km$^2$, comprises 1600 units, which are arranged on a triangular grid with 1500 m spacing. It samples the electromagnetic and muonic components of extensive air showers with a duty cycle of nearly 
100\%. Each water-Cherenkov unit is a 1.2 m depth, 10 m$^2$ area, detector containing 12000 liters of ultra-pure water. The water volume is viewed by three 9'' photomultiplier tubes (PMTs). Two signals (from the anode and from the amplified dynode) from each of 
PMTs are digitized by 40 MHz 10-bit Flash Analog to Digital Converters (FADCs). The recorded signals are calibrated in units of the signal produced by a muon traversing the water vertically. The unit is termed the ``Vertical Equivalent Muon'' or VEM 
\cite{Bertou:2006}. The shower-trigger requirement is based on a 3-fold coincidence, satisfied when a triangle of neighboring stations is triggered \cite{Abraham:2010zz}. These triggers result in the recording of $19.2\mu$s (in 768 bins) of data from each of the six FADCs in each triggered station. In the present analysis, that relies on the use of the \emph{risetime} of the signals (see section~\ref{sec:AnalysisTitle}), we use FADC traces from stations in events that are well-confined within the array, that is, the largest signal station is surrounded by 6 working stations, so that an accurate reconstruction is ensured. 
For these events, we determine the arrival directions of the primary cosmic rays from the relative arrival times of the shower front in the triggered stations. The 
angular resolution is 0.9$\degree$ for energies above $3 \times 10^{18}$ eV \cite{Bonifazi:2009}. The estimator of the primary energy is the reconstructed signal at 1000 m from the shower core, $S(1000)$. This is determined, together with the core position, through 
a fit of the recorded signals (converted to units of VEM after integration of the FADC traces) to a lateral distribution function that describes the average rate fall-off of the signal as a function of the distance from the shower core. For $S(1000) > 17$ VEM 
(corresponding to primary energy around $3 \times 10^{18}$ eV) the core location is determined with an uncertainty of 50 m, which is independent of the shower geometry \cite{Aab:2015zoa,Bonifazi:ICRC05}. 
The accuracy of the determination of $S(1000)$ is 12\% (3\%) at $3 \times 10^{18}$ ($10^{19}$) eV \cite{Ave:2007wf}. 

The conversion from this estimator to energy is derived through the use of a subset of showers that trigger the fluorescence detector and the surface array independently (``hybrid'' events). The statistical uncertainty in the energy determination is about 16\% 
(12\%) for energies above $3 \times 10^{18}$ ($10^{19}$) eV. The absolute energy scale, determined by the FD, has a systematic uncertainty of 14\% \cite{Verzi:2013}. In the following we use events for which the zenith angle is less than 62$\degree$ and the energy 
is above $3 \times 10^{18}$ eV: the efficiency of detection in such cases is 100\%. 

The fluorescence detector consists of 27 optical telescopes that overlook the array \cite{Abraham:2009pm,Mathes:2011}. On clear moonless nights, these are used to observe the longitudinal development of showers by detecting the fluorescence light produced in 
the atmosphere by charged particles along the shower trajectory. The duty cycle of the FD is $\sim$ 13\%. In hybrid events, by using the time constraint from the SD, the shower geometry can be determined from the arrival times at the FD and SD with an angular 
uncertainty of 0.6$\degree$. With the help of information from atmospheric monitoring devices \cite{Abraham:2010pf} the light collected by the telescopes is corrected for the atmospheric attenuation between the shower and the detector. Finally, from the shower 
geometry the longitudinal shower profile is reconstructed from the light recorded by the FD \cite{Aab:2014kda,Abraham:2010mj,Abraham:2010yv}. The $\Xmax$ value and the energy are determined by fitting the reconstructed longitudinal profile with a Gaisser-Hillas 
function \cite{Gaisser:1997bu}. The resolution of $\Xmax$ is around 20 g\,cm$^{-2}$ in the energy range of interest for this work.
%
 \section{\label{sec:Method} Concept of azimuthal asymmetry in the risetime}

The water-Cherenkov detectors are used to measure the signal size and the spread in arrival times of the signals produced by the different components of an extensive air shower. An approach originally used to analyze the data of the Haverah Park detector \cite{Watson:1974tf} showed 
that observables related to time-spread have sensitivity to the mass of the primary particle. In composition studies, the risetime, $\rt$, is usually employed to characterize the recorded signal. It is defined as the time of increase from 
10\% to 50\% of the total integrated signal. The average risetime is a function of the distance to the axis of the shower and of the zenith angle of that shower. In individual events it is necessary to take account of the time at which each detector 
is struck. Note that detectors that are hit later will register the shower after it has passed through additional atmosphere, and the particles detected, in particular the muons, will in general come from a smaller angle to the shower axis. To describe this we introduce the 
concept of ``early'' and ``late'' detectors (see Fig.~\ref{fig:CoordinateSystem}). We classify as ``early'' those detectors that record the passage of the shower front first. With our convention these correspond to detectors with polar angles $\mid\zeta\mid \,< \pi/2$ with respect to the direction of the shower axis projected on to the ground. Detectors in the $\mid\zeta\mid \,> \pi/2$ region are dubbed ``late''. 

The top two panels of Fig.~\ref{fig:event_stations} show the recorded signals for a nearly vertical event in an early station 
(left) and a late station (right) (the reconstructed zenith angle is 15.7$\degree$). The FADC traces can, to a good approximation, be considered equal in amplitude and time-spread. The bottom panels of Fig.~\ref{fig:event_stations} show two typical FADC signals recorded 
for an event with a reconstructed energy of 7.7 EeV and a zenith angle of 52$\degree$ (early and late as above). In this event, although both detectors are located at similar distances from the shower core, the traces are strikingly different, both in magnitude and time structure. We 
observed this asymmetric behavior (in total signal and time-spread) for the first time in the FADC traces recorded with the detectors of the Engineering Array constructed for the Observatory \cite{Dova:2003rz}.

To appreciate the origin of the asymmetries, we have to understand the behavior of the copious number of muons and electromagnetic particles that reach the ground. For a vertical shower of 10 EeV a signal of $\sim$ 50 VEM is recorded at 1000 m from the shower axis. About 50\% of the total signal is due to muons sufficiently energetic to traverse the detector without stopping. Electrons are a factor 10, and photons a factor 100, more numerous than muons. They make up the other 50\% of the total signal 
and, as they have average energies of $\sim$ 10 MeV \cite{Kellermann:1970iv}, are largely absorbed in the 3.2 radiation lengths of water in the station. The ratio of the muon to electromagnetic signal changes with distance and zenith angle in a manner that is known from dedicated 
measurements made at several of the early ground-detector arrays.  Greisen \cite{Greisen:1960wc} was the first to point out that attenuation of shower particles in the atmosphere would lead to a loss of circular symmetry in the signal intensities recorded by a 
detector at a single atmospheric depth. Experimental evidence of this effect was obtained by England \cite{England:1984} using data from Haverah Park. Regarding the risetime of the signals, Linsley and Scarsi \cite{Linsley:1962kq} demonstrated that the thickness of 
the disc of particles in the shower increased from a few meters near the axis to several hundreds of meters at large distances. Using Haverah Park data, a study showed that the spread of the arrival time distribution was decreased by attenuation across the 
array \cite{Baxter:1968}.

The observed azimuthal asymmetry is due to two effects. On the one hand, a contribution comes from the quenching of the electromagnetic signal. Since the particles that reach late detectors traverse longer atmospheric paths, we expect a bigger attenuation of electrons and photons as compared to early detectors. On the other hand, there are also contributions to the asymmetry from geometrical effects. In this case, not only is the electromagnetic component important, but muons also play a role. The cylindrical design of the the detectors affords longer possible paths within the detector at larger zenith angles, thus increasing the signal strength from muons and compensating somewhat for the reduced numbers of electrons and photons. The angular distributions of detected muons are different for higher zenith angle showers, as late detectors record more muons emitted closer to the shower axis. Geometrical effects predominate at small zenith angles, while for showers with $\theta > 30\degree$ attenuation effects are the main contribution.

\begin{figure}[t]
    \centerline{\includegraphics[width=0.55\textwidth]{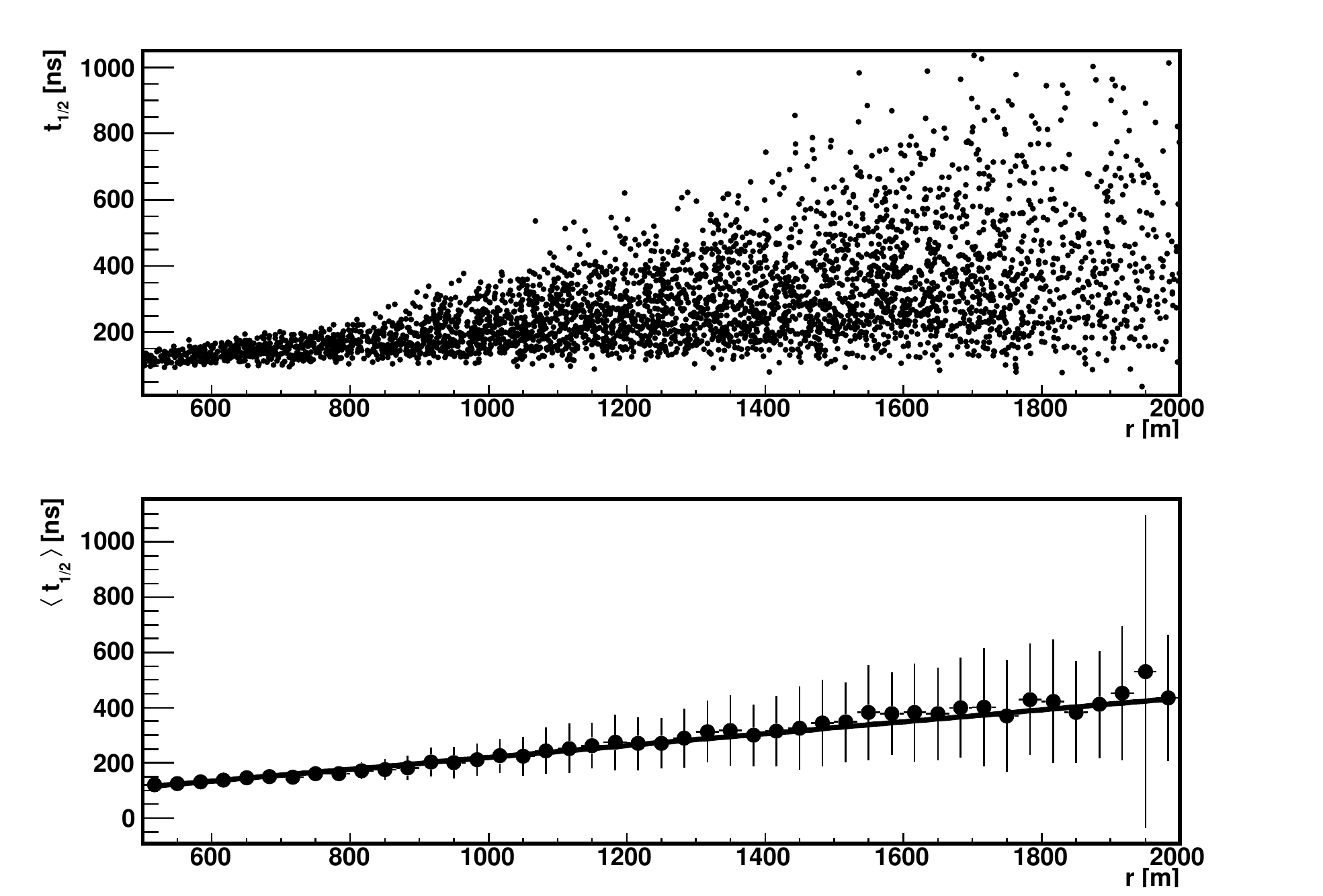}}
    \caption{{Example of risetime vs core distance for stations in events between energies 10$^{19.2} - 10^{19.6}$ eV and zenith angle 42$\degree - 48\degree$. Top: scatter distribution of the risetime values for individual stations. Bottom: bin-by-bin averages of the risetime. 
	Vertical bars represent the root-mean-square of the corresponding distributions.
	}}
    \label{fig:rt_vs_r}
\end{figure}

As already mentioned, it is known that the risetime has a dependence with respect to the distance of the detector to the core of the shower in the plane of the shower front, $r$ \cite{Watson:1974tf}. Fig.~\ref{fig:rt_vs_r} shows that $\rt$ is an increasing function 
of distance. For the range of distances selected in this work, this function can be approximated to first order as a straight line. But the risetime is not the only observable showing a distance dependence. Based on the previous considerations we expect that the 
asymmetry will also show a dependence on core distance. For measurements close to the shower axis, the path difference between late and early detectors is not large and therefore we do not expect a sizeable asymmetry. It becomes more evident as the distance increases. 

The azimuthal asymmetry of the risetime must also depend on the zenith angle. As suggested earlier in Fig.~\ref{fig:event_stations}, no asymmetry is expected for vertical showers but it is expected to grow as the zenith angle increases (and therefore differences in atmospheric paths become larger for a given set of 
triggered detectors). However this trend reaches a point where it does not hold for more horizontal events. For these the electromagnetic signal is quenched due to the longer atmospheric path travelled and the particles in the showers are dominantly muons. This 
translates into a reduction of the asymmetry as $\theta$ approaches 90$\degree$. As discussed in \cite{Dova:2009fk,DiegoGP:Th}, for a given energy $E$, the azimuthal asymmetry dependence upon $\sec\theta$ shows a correlation with the average longitudinal 
development of the shower. Hence the time asymmetry is sensitive to the average mass of the primary cosmic ray. 

\begin{figure}[t]
    \includegraphics[width=0.50\textwidth]{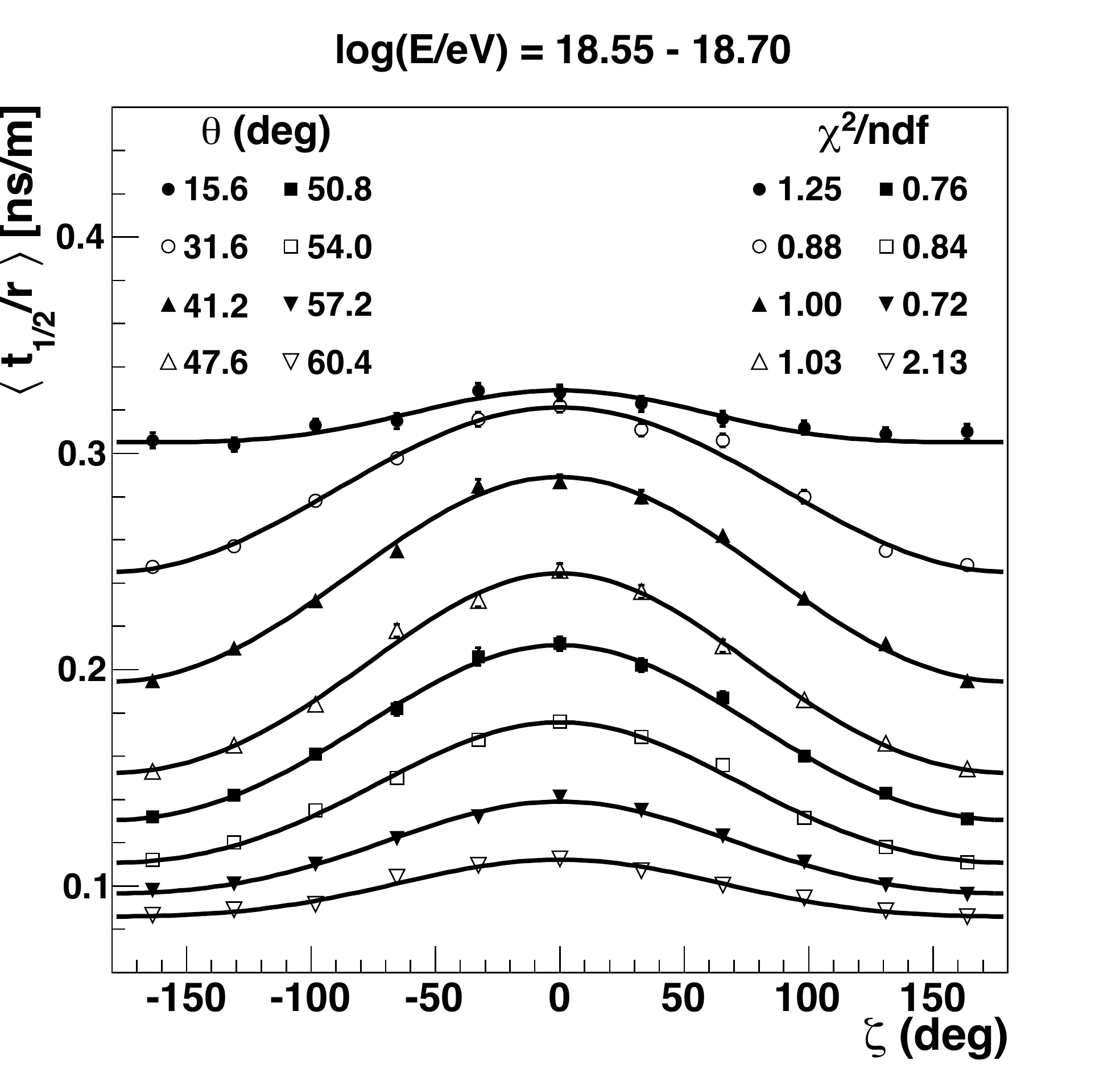}
    \includegraphics[width=0.50\textwidth]{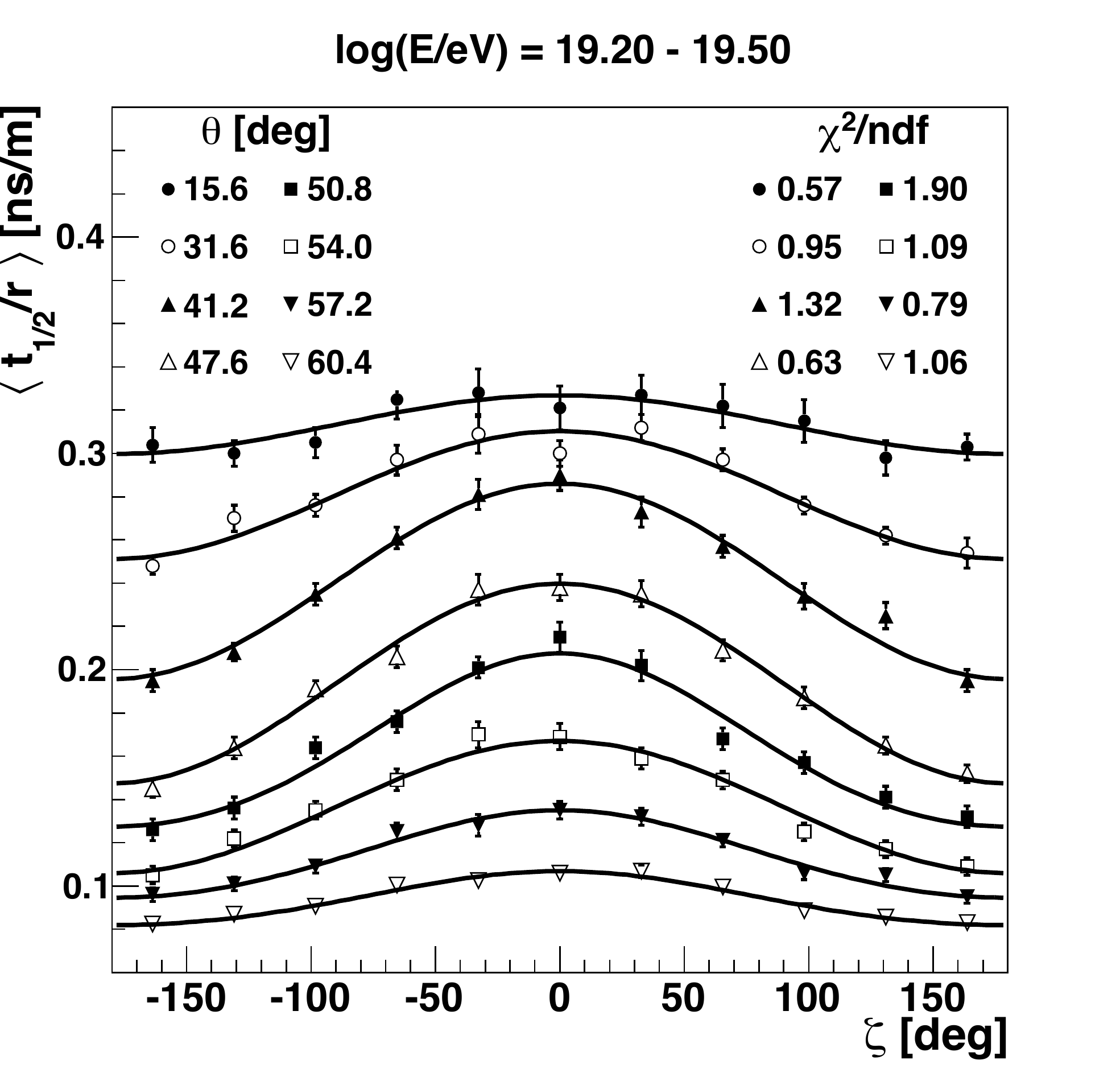}
    \caption{{Dependence of $\meanrt$ on the polar angle $\zeta$ in the shower plane for primary energy $\log{\mathrm(E}$/eV) = 18.55 $-$ 18.70 (top) and 19.20 $-$ 19.50 (bottom) at different zenith angles bands. Each data point represents an average (with the 
corresponding uncertainty) over all stations surviving the selection criteria (see text).}}
    \label{fig:RTvsZeta}
\end{figure}


\section{\label{sec:AnalysisTitle} Azimuthal asymmetry using Auger data}

\subsection{\label{subsec:Analysis} The analysis}

The first step in the analysis is the measure of the $\rt$ value in each detector. We use the events collected with the surface array of the Pierre Auger Observatory from January 2004 to October 2014. We consider only the FADC traces of the events that pass the selection criteria described in section~\ref{sec:Observatory}. Those traces allow us to compute the average of the risetimes of active PMTs in every station.
Quality cuts on data have been applied, based on core distance and total recorded signal. We have required that the recorded signal is larger than 10 VEM, above which level the probability of single detector triggering is about 100$\%$ \cite{Abraham:2010zz}. With 
respect to core position, detectors used for the analysis were required to be further than 500 m from the core of the shower to avoid signal saturation effects that prevent an accurate measurement of $\rt$ (signals saturate at average values of about 800 VEM 
depending on the PMT gains and the risetime of the signal). The uncertainty of the measured risetimes is estimated comparing measurements of the same parameter from multiple observations: twins (stations separated by 11 m) or stations belonging to the same event with core distance difference 
smaller than 100 m \cite{Wileman:Th,Smith:Th}. It is required that the water-Cherenkov detectors are within 2 km of the core: this is a fiducial cut to exclude stations with high uncertainties in the reconstructed risetimes.
After application of the station selection criteria, a total of 191534 FADC signals from 54584 events remain. 

The second step consists in measuring the azimuthal asymmetry of the risetime distributions as a function of the polar angle, for fixed energies and zenith angles. This measurement cannot be done on a shower-by-shower basis because it is not possible to sample the 
whole range of the polar angle, from early to late regions, in a single event. Thus, a statistical approach is applied to characterize the azimuthal asymmetry of the risetime as a function of the polar angle, using all the stations from the events at a given energy 
and zenith angle.

The risetime grows with the core distance $r$, and in a first approximation, follows a linear behavior in the range of distances considered in the present analysis as was seen in Fig.~\ref{fig:rt_vs_r}. The variable used to study the azimuthal asymmetry is 
$t_{1/2}/r$. This quantity is chosen since an average value using all stations at different core distances, allowing an increase in the number of events used, can be computed and thus the asymmetry information from the whole $r$ range can be used in the analysis. To derive the 
behavior of the asymmetry vs polar angle we thus use the value $\meanrt$  averaged over all stations in all events that fulfill the criteria described above in defined bins of energy and angle.

As an example, we show in Fig.~\ref{fig:RTvsZeta} the values of $\meanrt$ vs $\zeta$ for eight zenith angles and for a narrow interval of energy centered on $4.2 \times 10^{18}$ eV (top panel) and on $2.2 \times 10^{19}$ eV (bottom panel). For each zenith-angle band 
the data are fitted to the function $\meanrt = a + b\cos\zeta + c\cos^2\zeta$. The asymmetry with respect to $\zeta$ is evident and the ratio $b/(a+c)$, the so-called asymmetry factor, is used to give a measure of the asymmetry. In Fig.~\ref{fig:RTvsZeta} 
results for a wide range of zenith angles are shown although the analysis has been restricted to the interval 30$\degree - 62\degree$.

\begin{figure}[t!]
  \begin{center}
    \includegraphics[width=0.5\textwidth]{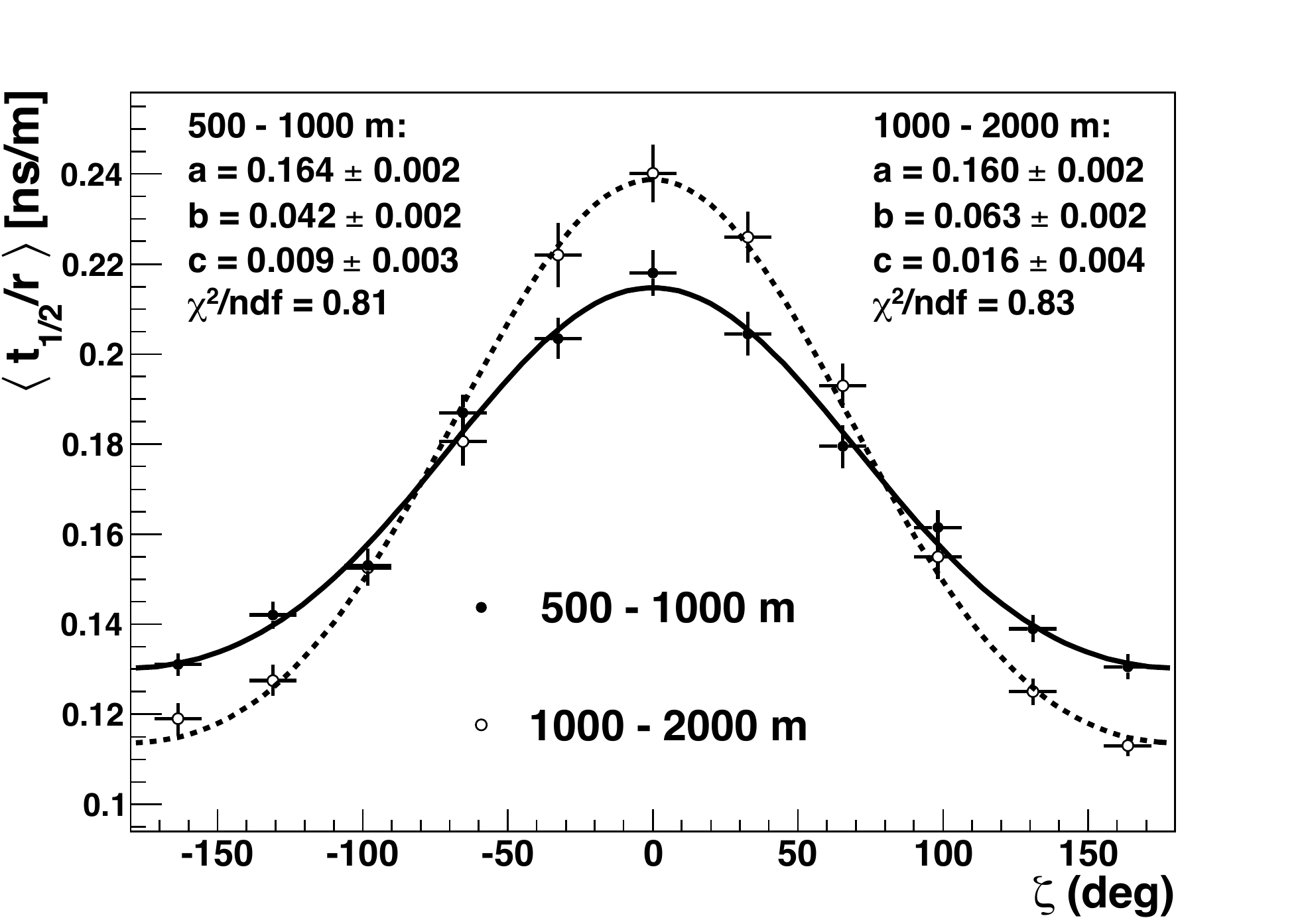}
    \caption{Dependence of $\meanrt$ on $\zeta$ for two chosen core distance intervals for data. Results of the fitted parameters (see text) are shown for each core distance interval.}
    \label{fig:plot_rDependence}
  \end{center}
\end{figure}

As mentioned before the asymmetry depends on the distance to the core position. To take that into account the analysis has been carried out independently for two $r$-intervals, i.e., 500 $-$ 1000 m and 1000 $-$ 2000 m. This selection leads to a total of 102123 FADC 
signals from stations passing the cuts for the 500 $-$ 1000 m interval, and 89411 FADC signals for the 1000 $-$ 2000 m interval. As an example in Fig.~\ref{fig:plot_rDependence} $\langle t_{1/2}/r \rangle$ vs $\zeta$ is displayed for both core 
distance intervals for showers with $\log(E$/eV) = 19.1 and $\theta$ = 51$\degree$. The smaller asymmetry amplitude of the 500 $-$ 1000 m is evident. This is due to the fact that, close to the core there is a smaller difference in the paths travelled by the particles.

\begin{figure*}[t!] 
    \includegraphics[width=0.40\textwidth]{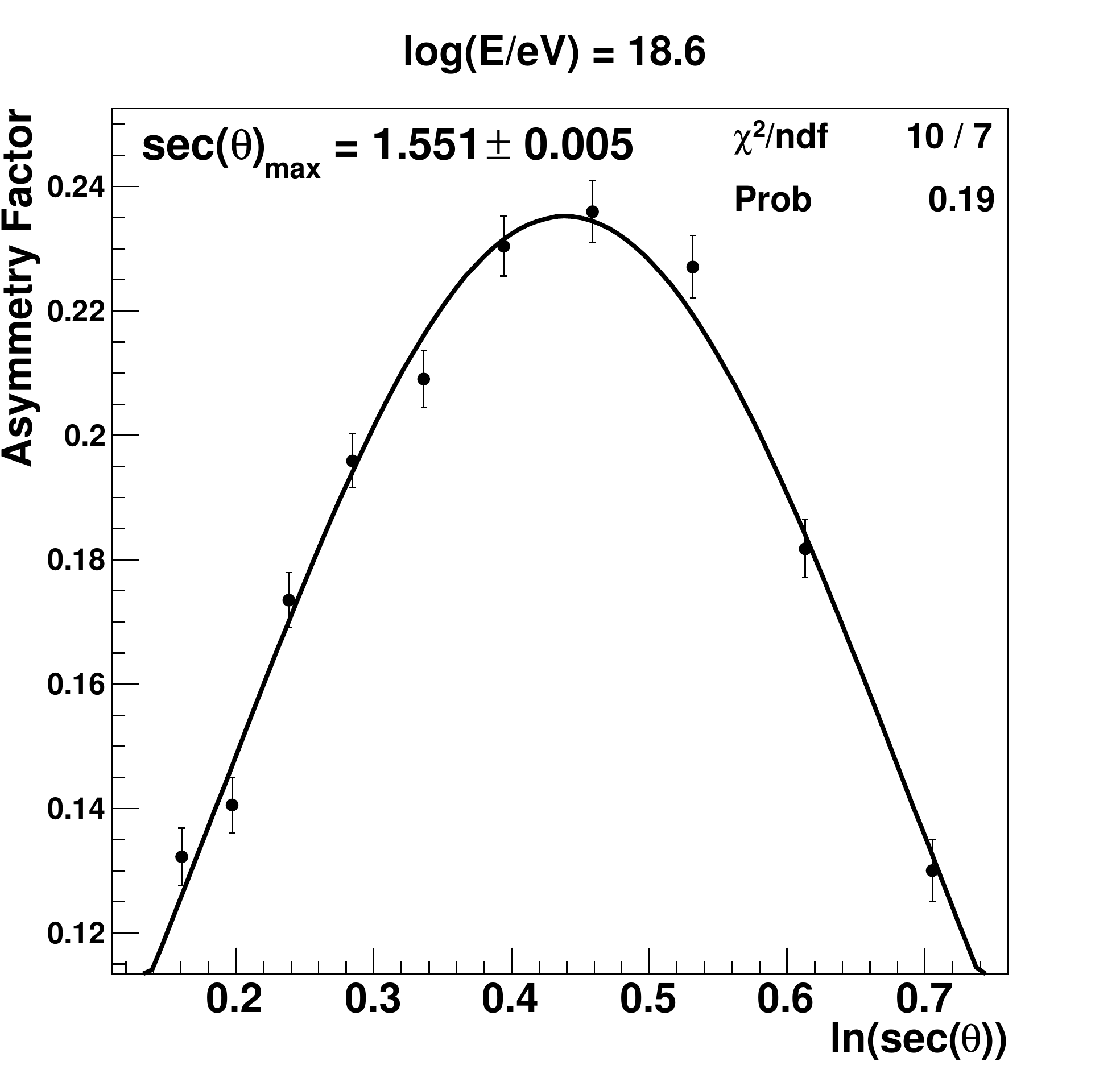} 
    \includegraphics[width=0.40\textwidth]{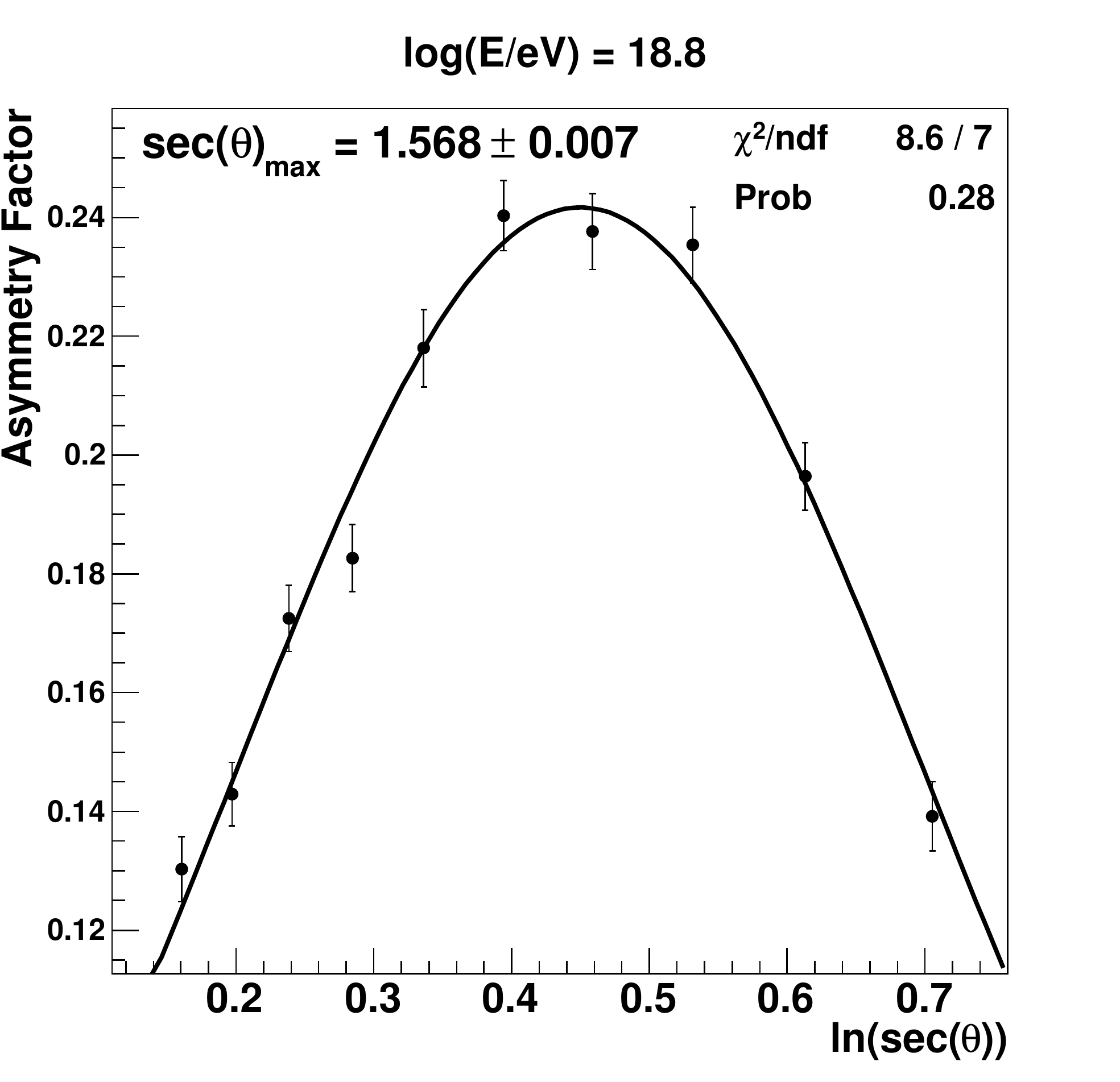}
    \includegraphics[width=0.40\textwidth]{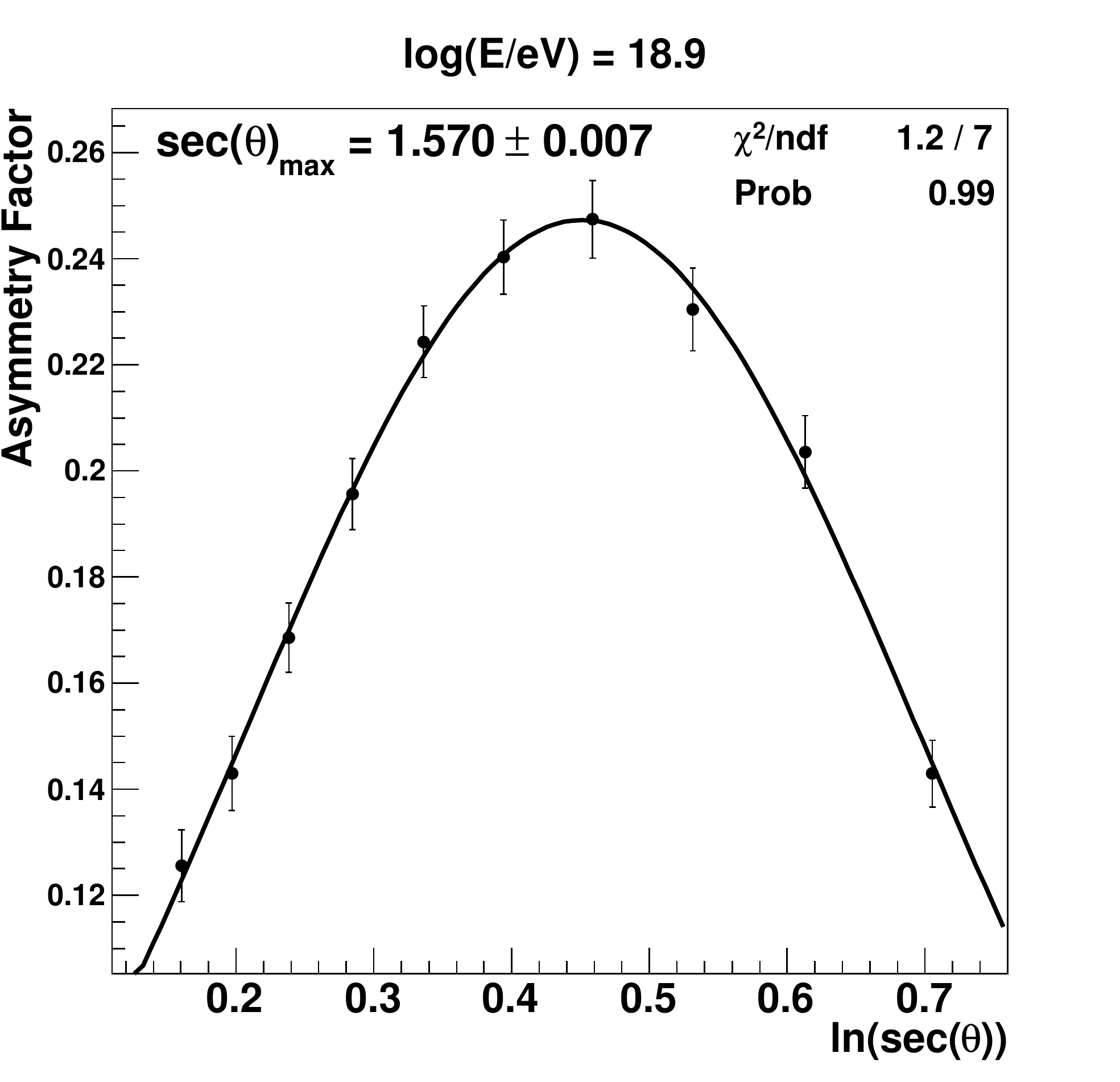} 
    \includegraphics[width=0.40\textwidth]{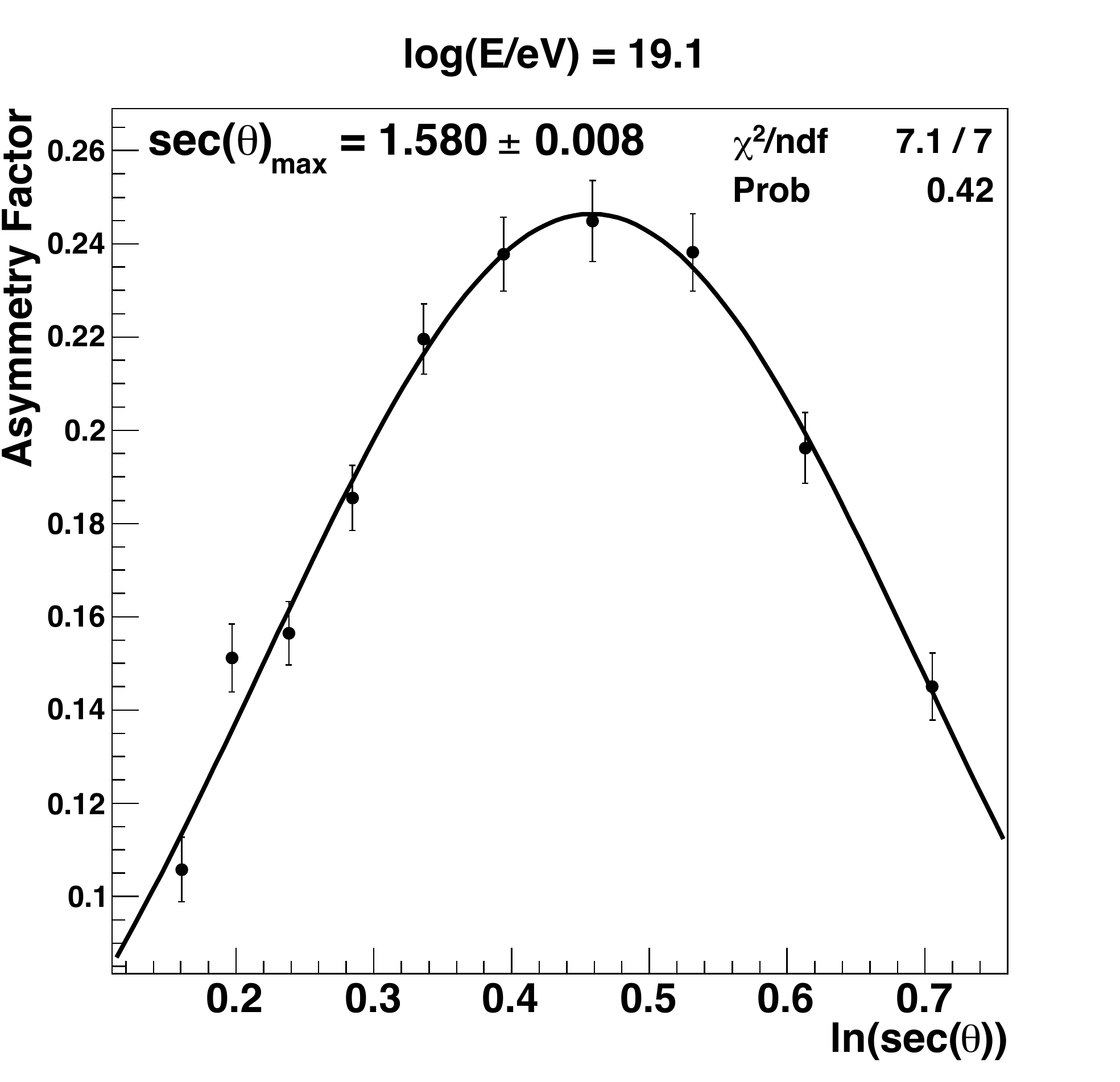}
    \includegraphics[width=0.40\textwidth]{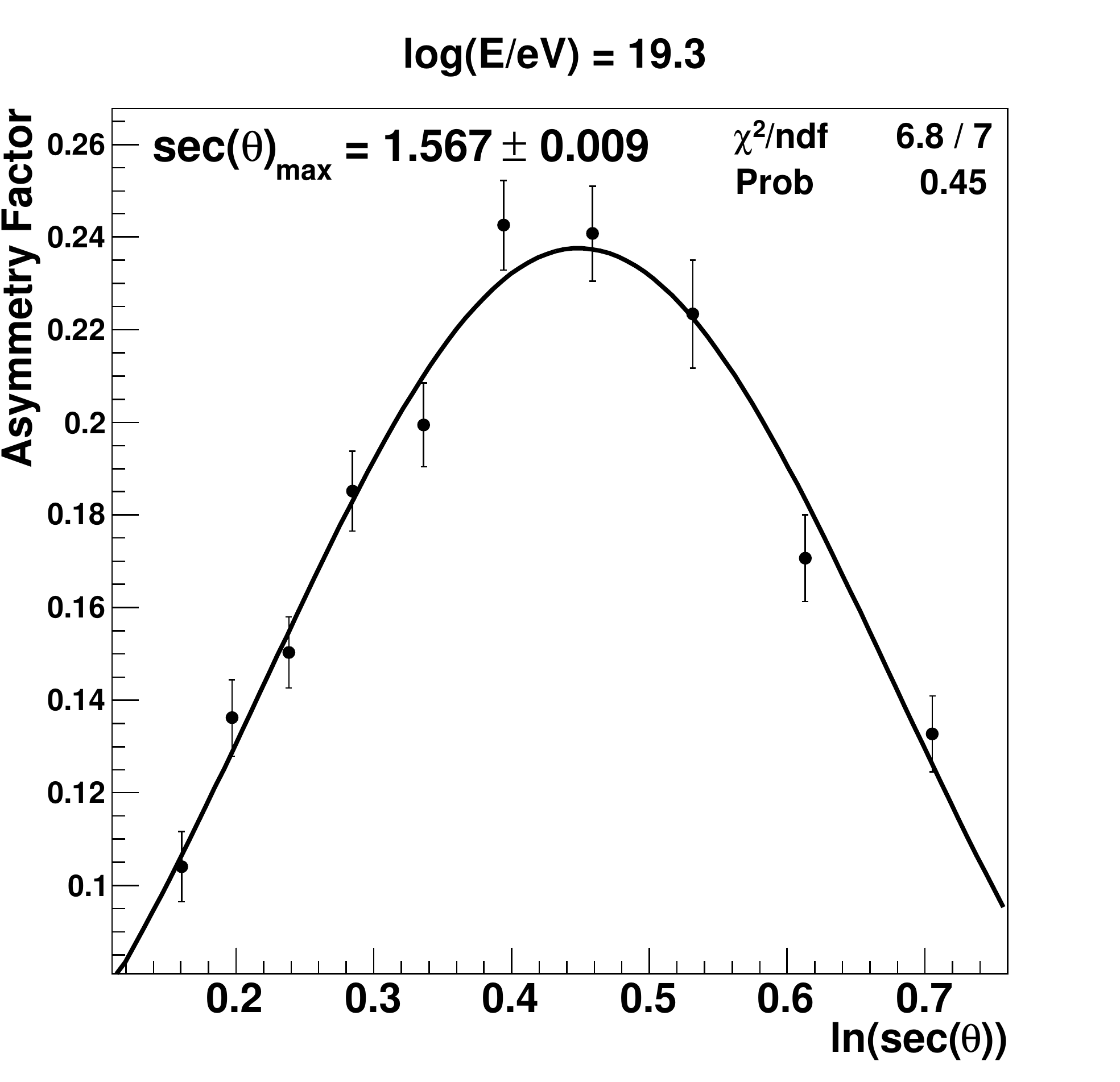}
    \includegraphics[width=0.40\textwidth]{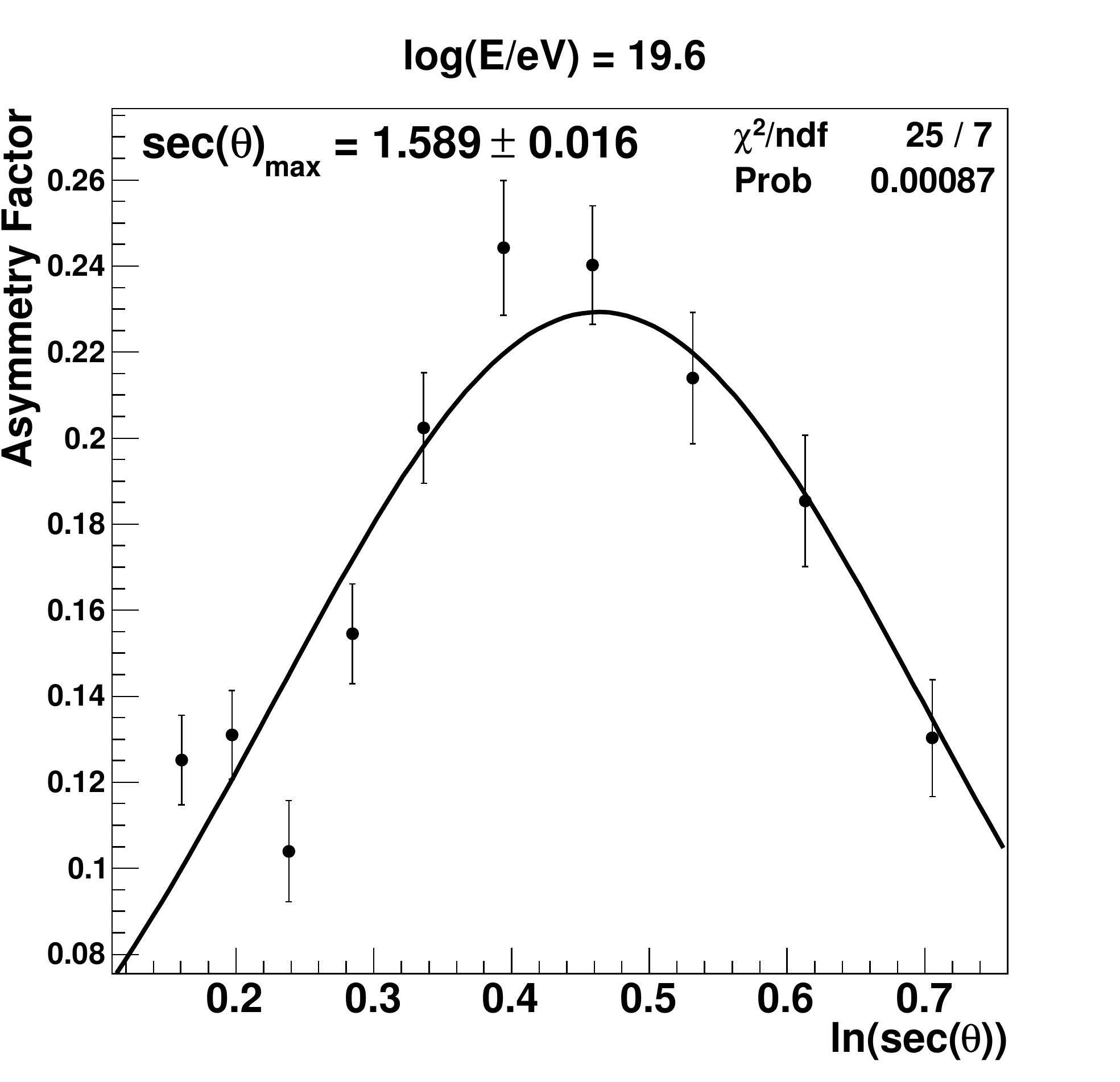}
\caption{Asymmetry longitudinal development in bins of $\log{\mathrm(E/eV)}$ at the interval 500 $-$ 1000 m . From left to right and top to bottom: $18.55 - 18.70, 18.70 - 18.85, 18.85 - 19.00, 19.00 - 19.20, 19.20 - 19.50$ and above $19.50.$} 
\label{fig:long_500_1000}
\end{figure*}
\begin{figure*}[t!]
  \includegraphics[width=0.40\textwidth]{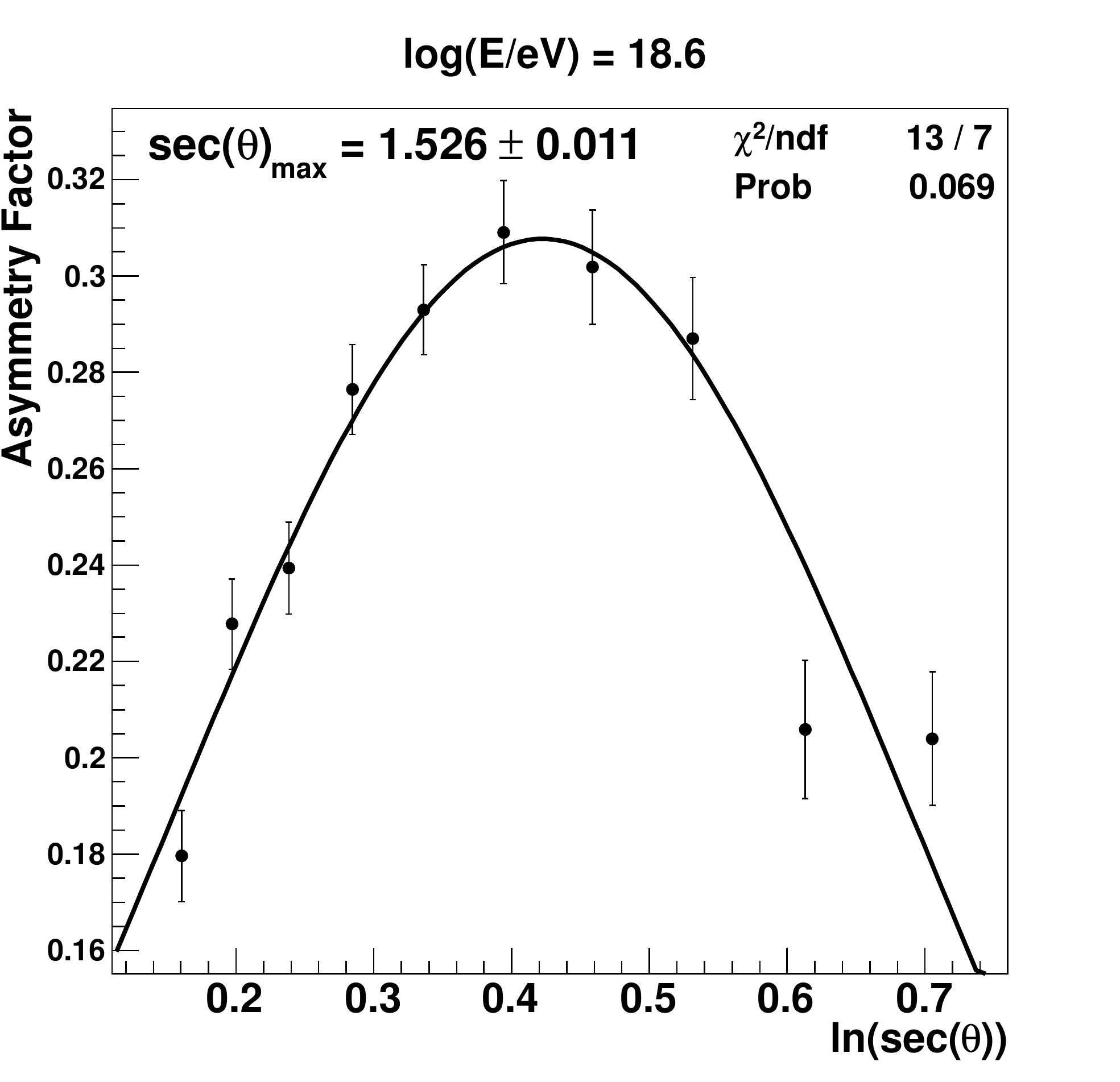}
  \includegraphics[width=0.40\textwidth]{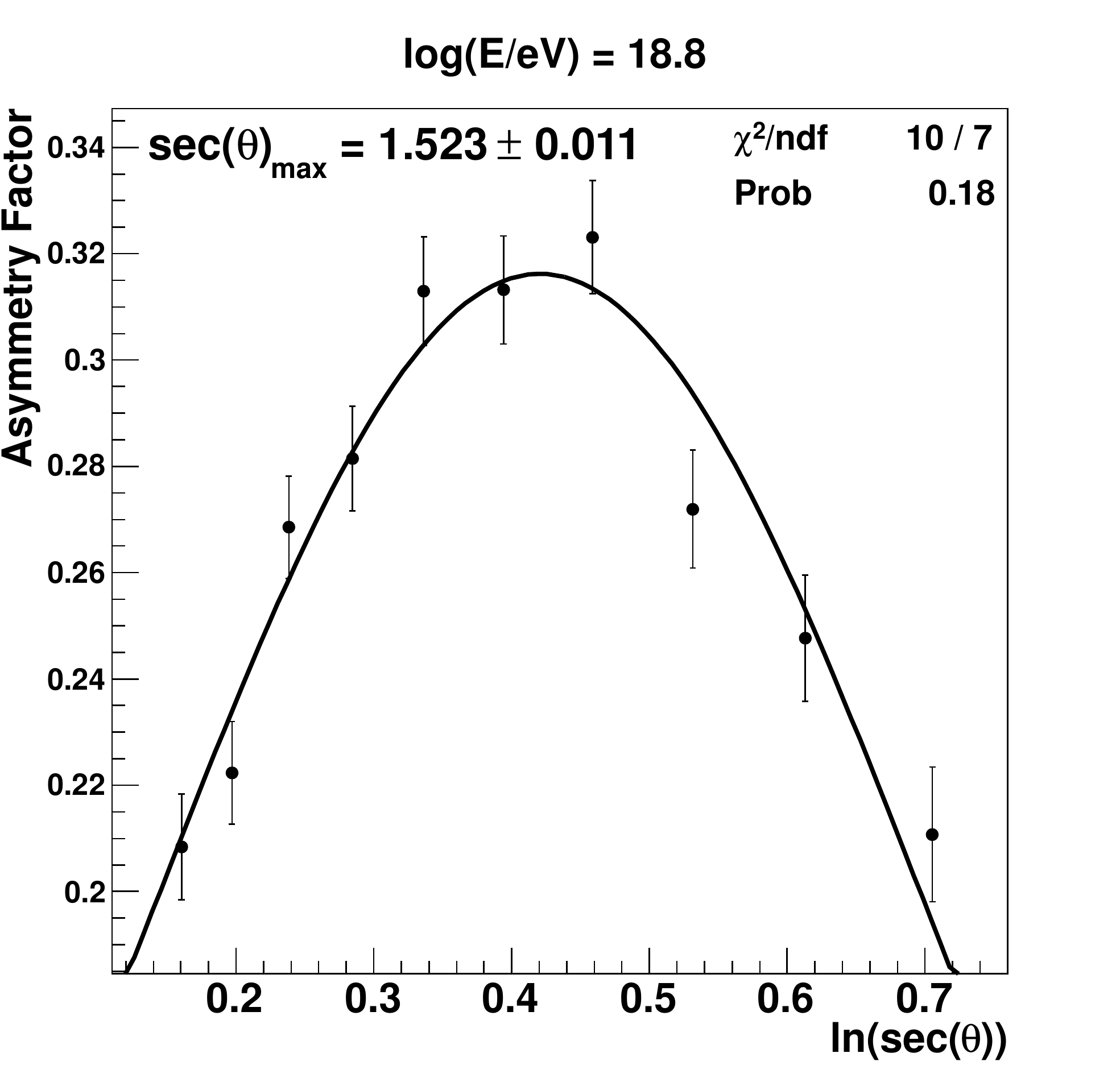}
  \includegraphics[width=0.40\textwidth]{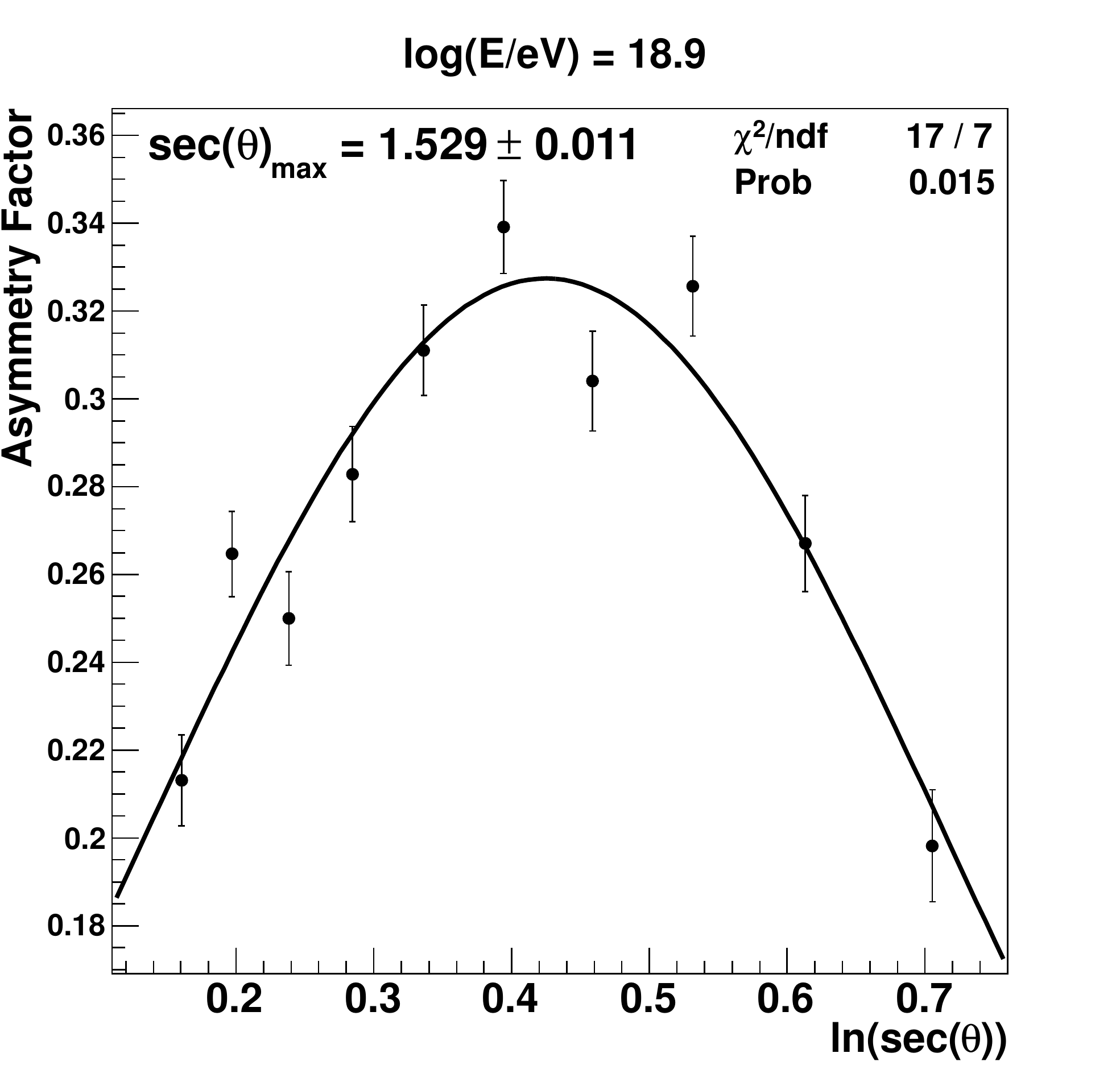}
  \includegraphics[width=0.40\textwidth]{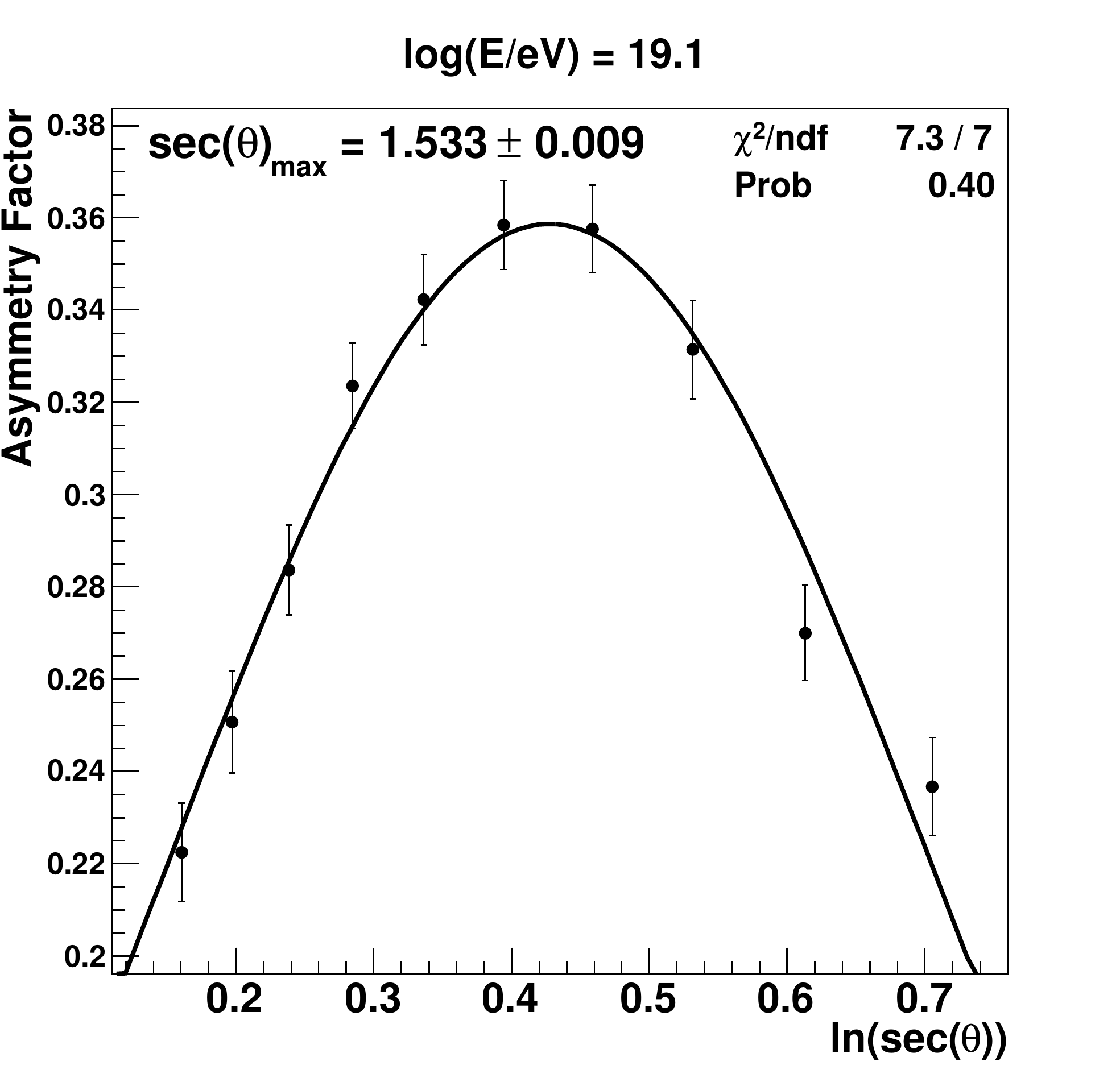}
  \includegraphics[width=0.40\textwidth]{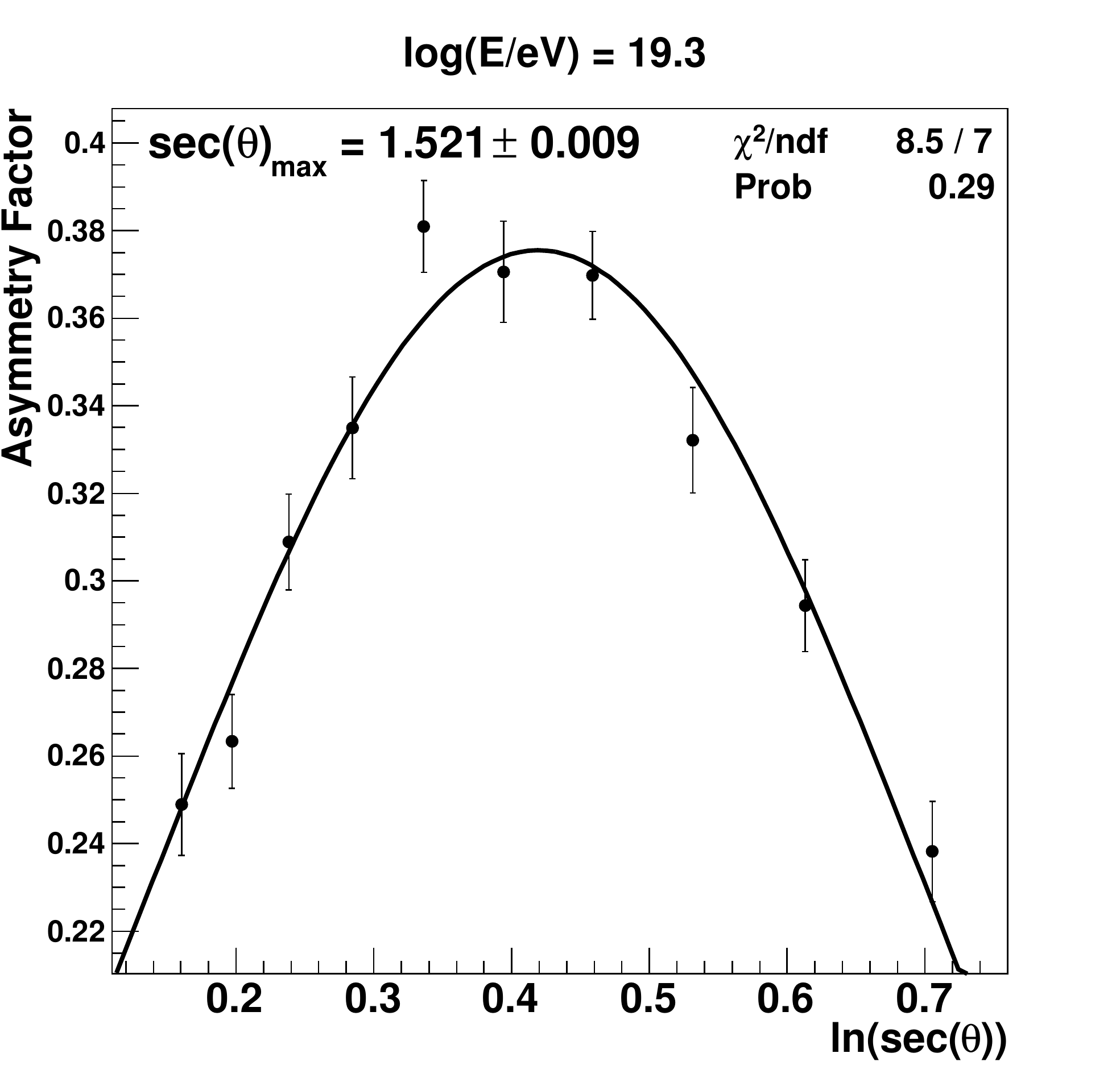}
  \includegraphics[width=0.40\textwidth]{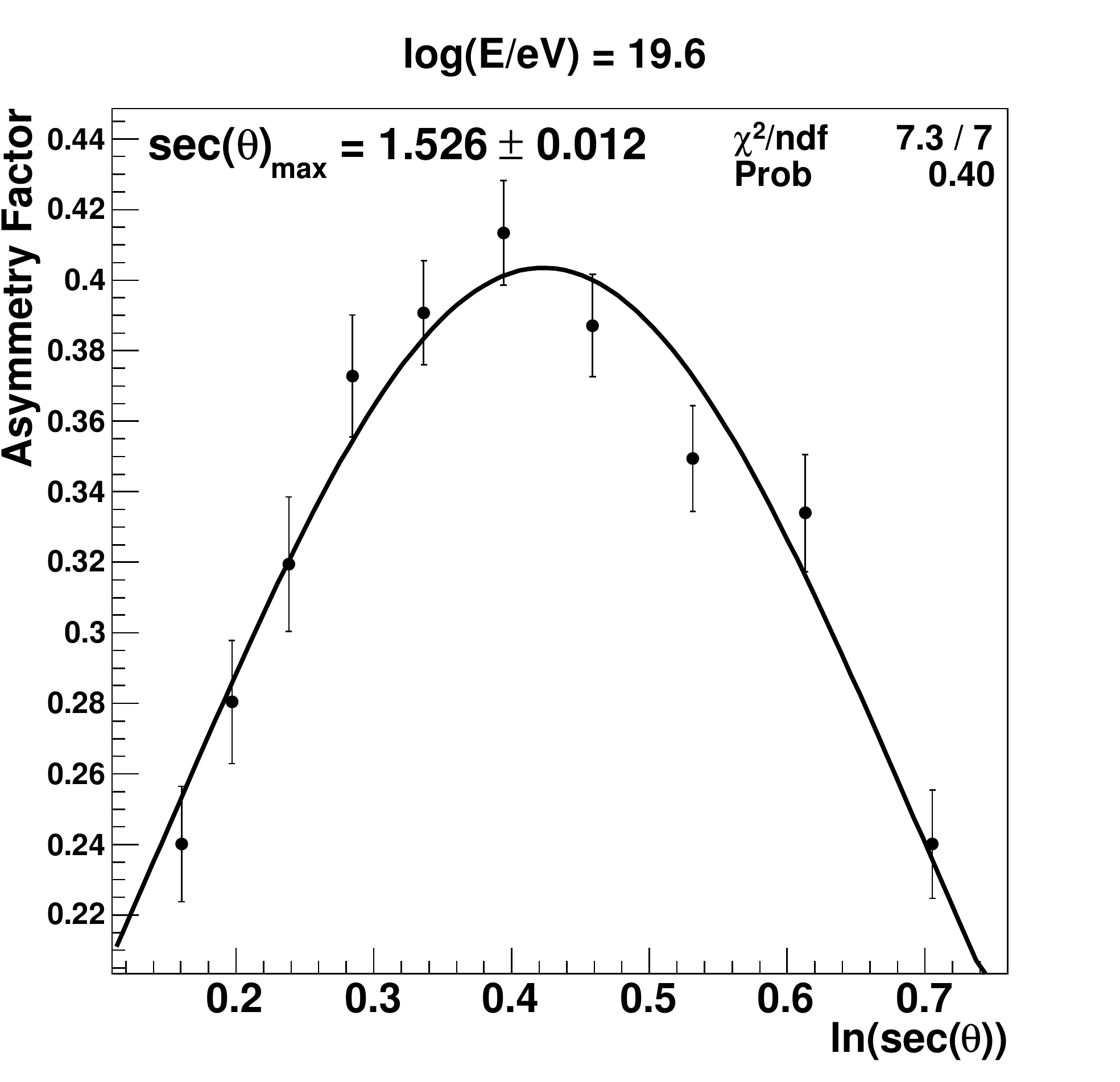}
  \caption{Asymmetry longitudinal development in bins of $\log{\mathrm(E/eV)}$ at the interval 1000 $-$ 2000 m. From left to right and top to bottom: $18.55 - 18.70, 18.70 - 18.85, 18.85 - 19.00, 19.00 - 19.20, 19.20 - 19.50$ and above $19.50.$}
  \label{fig:long_1000_2000}
\end{figure*}

The next step of the analysis is the study of the behavior of the asymmetry factor as a function of atmospheric depth, measured by $\sec \theta$. In Figs.~\ref{fig:long_500_1000} and \ref{fig:long_1000_2000}, $b/(a+c)$ has been plotted versus $\ln(\sec \theta)$ 
for six energy bins and for both core distance intervals. It is evident that for a given primary energy, the azimuthal asymmetry depends on zenith angle of the primary cosmic ray. 

For each energy interval, the dependence of the asymmetry parameter on $\ln(\sec \theta)$  is fitted using a Gaussian function. From this fit we can determine the value of $\ln(\sec \theta)$ for which the asymmetry parameter maximizes, and the corresponding 
$\secMax$ value will be used as the observable to describe the longitudinal evolution of the shower and thus with capability for the analysis of the mass composition. 

The dependence of the asymmetry on the core distance leads to a dependence of $\secMax$ on the $r$ interval of the station sample used in the analysis, as we can see in Figs.~\ref{fig:long_500_1000} and \ref{fig:long_1000_2000}. Apart from geometrical effects this can be understood as follows. Closer to the shower core  (500 $-$ 1000 m) there are electrons (and photons) with higher energies than those at larger distances, thus the electromagnetic cascade dies out deeper in the atmosphere than it does at larger distances. Hence, the symmetric influence of muons shows up deeper in the atmosphere for 500 $-$ 1000 m than it does for 1000 $-$ 2000 m. Therefore, selecting stations close to the core leads to systematically larger $\secMax$ values as expected since closer to the core the asymmetry is smaller, and thus, the zenith angle at which the muon component starts to dominate (and the asymmetry starts to decrease) is higher.
\subsection {\label{subsec:systematic} Systematic Uncertainties}

The sources of  systematic uncertainties related to the precision with which the absolute value of  $\secMax$ can be measured are discussed in the following. Results are presented 
in units of $\secMax$ which has a typical value of $\sim$ 1.55, and summarized in Table~\ref{Table_sys}.

\vspace{0.5cm}
{\it Risetime uncertainties}. A source of systematic uncertainty is that from the determination of the risetime itself.
To evaluate the effect of this uncertainty, the risetime has been shifted randomly around a Gaussian distribution with standard deviation $\sigma$ given by the uncertainty in the measurement of the risetime as mentioned in section~\ref{subsec:Analysis}. 
A systematic uncertainty of $+0.0008$/$-0.0063$ is obtained for the 500 $-$ 1000 m interval and $+0.0032$/$-0.0076$ for the 1000 $-$ 2000 m interval.

{\it Risetime parametrization}. The use of different parametrizations in the dependency of the risetime with the distance to the core is another possible source of uncertainty in $\secMax$. The dependence of the results on the particular choice of function has 
been checked by replacing the linear function used in the analysis by a quadratic function. This implies a redefinition of the parameter, using then $\langle t_{1/2}$/($a$ + $b\,r$ + $c\,r^{2}$)$\rangle$ instead of $\langle t_{1/2}$/$r \rangle$.  
The estimated systematic uncertainties are $+0.0019$/$-0.0012$ for the interval 500 $-$ 1000 m and $+0.0031$/$-0.0005$ for the interval 1000 $-$ 2000 m.

{\it Selection efficiency}. To evaluate a potential bias of the results towards a particular nuclear composition, we produced Monte Carlo samples of mixed composition (25\% p $-$ 75\% Fe, 50\% p $-$ 50\% Fe and 75\% p $-$ 25\% Fe) with both hadronic models QGSJETII-04 and EPOS-LHC. 
The samples were analyzed and the results were compared with the known input composition. The maximum deviations correspond to the 50\%$-$50\% composition and are taken as a systematic uncertainty. The values are of $\pm$0.010 units for 
both core distance intervals and both hadronic models.

{\it Core position reconstruction}. The systematic uncertainty arising from the reconstruction of the shower core was determined by shifting in the late direction (see section~\ref{sec:Method}) the position of the core by 50 m, corresponding to the typical shift to the early regions in inclined showers due to the asymmetry in the signal intensity. The whole chain of analysis to obtain  the new values of the position of the maximum of the asymmetry was repeated. 
The systematic uncertainty in units of $\secMax$ are $+0.0005$/$-0.0001$ for the 500 $-$ 1000 m interval and $+0$/$-0.0056$ for the 1000 $-$ 2000 m interval.

{\it Energy scale}. The absolute energy calibration of the Observatory is affected by a total systematic uncertainty of 14$\%$ \cite{Verzi:2013}. To study the corresponding effect on $\secMax$,  the energy values assigned to each event were 
shifted by the corresponding percentage and the full chain of the analysis was repeated. The shift leads to an uncertainty of $+0.0078$/$-0.0095$ for the 500 $-$ 1000 m interval and $+0.0090$/$-0.0030$ in 
units of $\secMax$ for the 1000 $-$ 2000 m interval.

{\it Additional Cross-Checks}. The systematic uncertainties estimated above have been validated by performing numerous cross-checks on the stability of the results. The most significant studies are: i)  a potential dependence on 
$\secMax$ due to the selection cuts in the signal intensity was studied by shifting the upper and lower cuts in the signal size; ii) the effect of the cuts on the angular intervals of the sample was also studied by varying the angular limits of the nominal 
interval; iii) the lateral width of the shower (in particular of the electromagnetic component) depends on pressure and temperature. A possible bias affecting the risetime measurements and hence $\secMax$  was evaluated splitting the data into ``hot'' 
(summer and spring) and ``cold'' (winter and autumn) periods and repeating the whole analysis chain for each case. iv) possible effect of aging \cite{Aab:2015zoa,Sato:2011zze} of the SD detectors on the results were studied separating the data sample in two equal sets, ``old'' (Jan.2004 $-$ Jan.2011) and ``new'' (Jan.2011 $-$ Oct.2014).
The first i) and ii) studies yield a maximum variation of $\secMax$ of 0.0044 which is well within the systematic uncertainties. In the case of iii) and iv) differences are compatible with zero within the statistical uncertainties of each sample.

\begin{figure}[t]
  \begin{center}
        \centerline{\includegraphics[width=0.45\textwidth]{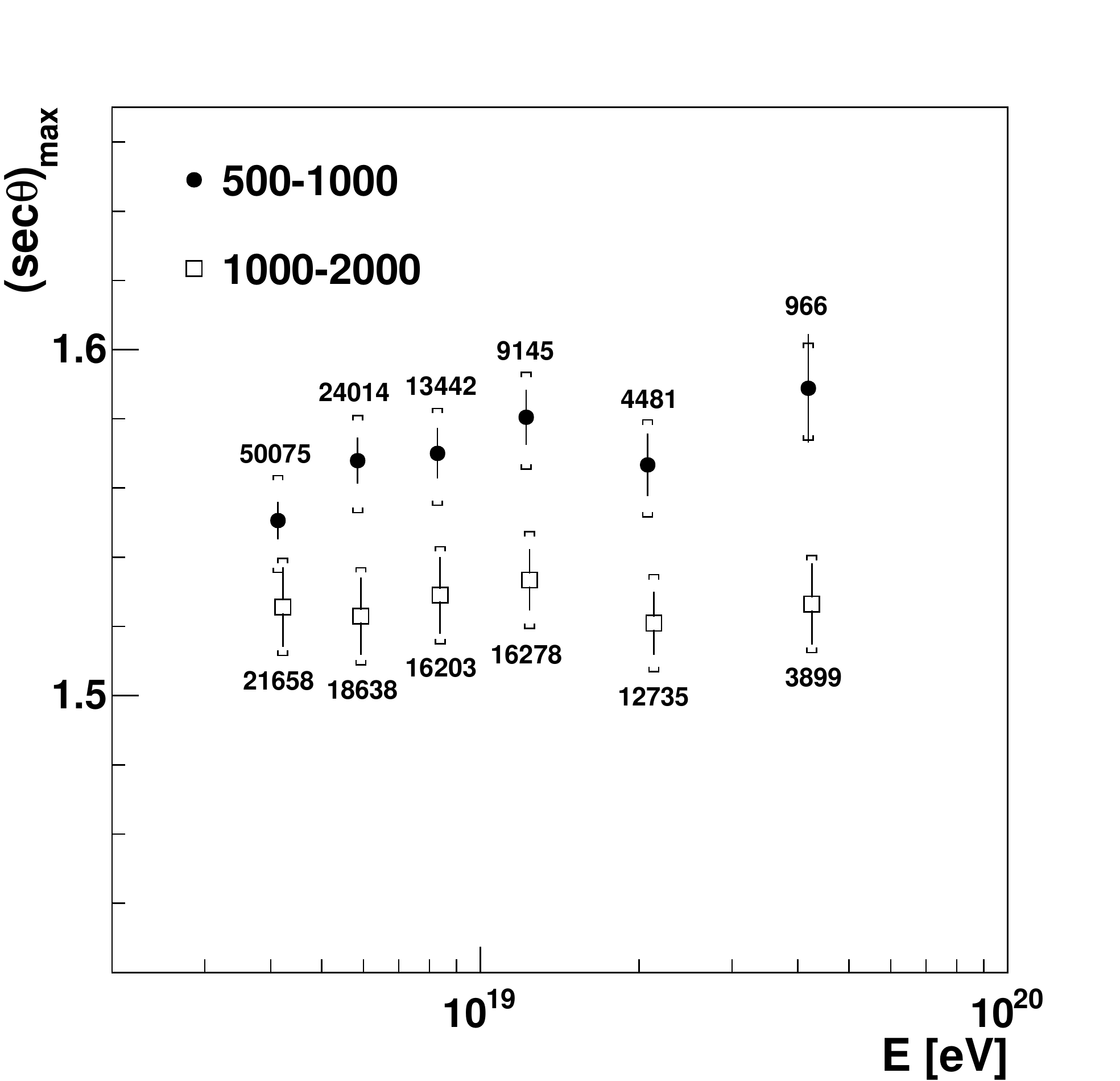}}
        \caption{Energy dependence of $\secMax$ for both intervals of core distance 500 $-$ 1000 m and 1000 $-$ 2000 m. Brackets represent the systematic uncertainty and the vertical lines the statistical uncertainties. The number of stations used for the analysis are indicated.}\label{plot_resultsintervals}
  \end{center}
\end{figure}

The overall systematic uncertainty (see Table~\ref{Table_sys}) in each radial interval amounts to $+0.013$/$-0.015$ for the 500 $-$ 1000 m interval, and $+0.014$/$-0.014$ for the 1000 $-$ 2000 m range. These values can be compared with the corresponding statistical uncertainties; for example, at a mean energy of $\log(E$/eV) = 19.1 and 500 $-$ 1000 m, $\secMax$ = 1.580 $\pm$ 0.008 (stat) $^{+0.013}_{-0.015}$ (sys), while for the 1000 $-$ 2000 m at the same energy the result is $\secMax$ = 1.533 $\pm$ 0.009 (stat) $^{+0.014}_{-0.014}$ (sys). Our analysis is therefore dominated by systematic uncertainties.

\begin{table}[hb]
  \begin{center}
    \begin{tabular}{ | l | c | c | c | c | }
    \hline
    {\bf Source of systematic} & \multicolumn{2} { | c | }{\bf 500 $-$ 1000 m} & \multicolumn{2} { | c | }{\bf 1000 $-$ 2000 m} \\ \hline
    Risetime uncertainties & $+0.0008$ & $-0.0063$ & $+0.0032$ & $-0.0076$ \\  \hline
    Risetime parametrization & $+0.0019$ & $-0.0012$ & $+0.0031$ & $-0.0005$ \\  \hline
    Selection efficiency & $+0.010$ & $-0.010 \; $ & $+0.010$ & $-0.010 \; $ \\  \hline
    Core position reconstruction & $+0.0005$ & $-0.0001$ & $+0$ & $-0.0056$ \\ \hline
    Energy scale & $+0.0078$ & $-0.0095$ & $+0.0090$ & $-0.0030$ \\  \hline
    \hline
    {\bf Total systematic value} & {\bf $+$0.013} & {\bf $-$0.015} & {\bf $+$0.014} & {\bf $-$0.014} \\
    \hline
    \end{tabular}
    \caption{Contributions to systematic uncertainty of $\secMax$ for all sources in both core distance intervals. Values are summed in quadrature to obtain the final systematic result.
    }\label{Table_sys}
  \end{center}
\end{table}

\section {\label{sec:results} Results}

Once the value of $\secMax$ for each energy bin has been obtained in each core distance interval, we can perform the final step of the asymmetry analysis, that is, the evaluation of the dependence of $\secMax$ on the primary energy. In Fig.~\ref{plot_resultsintervals} this result for both $r$ intervals is shown. 

To extract mass estimates from the measurements one must rely on the comparison with predictions made using current models of hadronic interactions extrapolated to these energies. For this purpose, a library of Monte Carlo events generated with the CORSIKA code \cite{Heck:1998vt} has been produced using the EPOS-LHC and QGSJETII-04 hadronic interaction models for two different primary species: proton and iron. 
A total of 77000 events (38500 of each primary) have been produced for each interaction model. The $\log(E$/eV) values ranged from 18.00 to 20.25 in bins of 0.25 with eleven discrete zenith angles between 18$\degree$ and 63$\degree$. 

Note that, in principle, the dependence of the $\secMax$ on $E$ with the radial interval shown in Fig.~\ref{plot_resultsintervals} should not limit the capability of the asymmetry method for mass analysis provided Monte Carlo simulations are able to correctly reproduce this dependence.

\begin{figure}[t!]
  \begin{center}
        \includegraphics[width=0.45\textwidth]{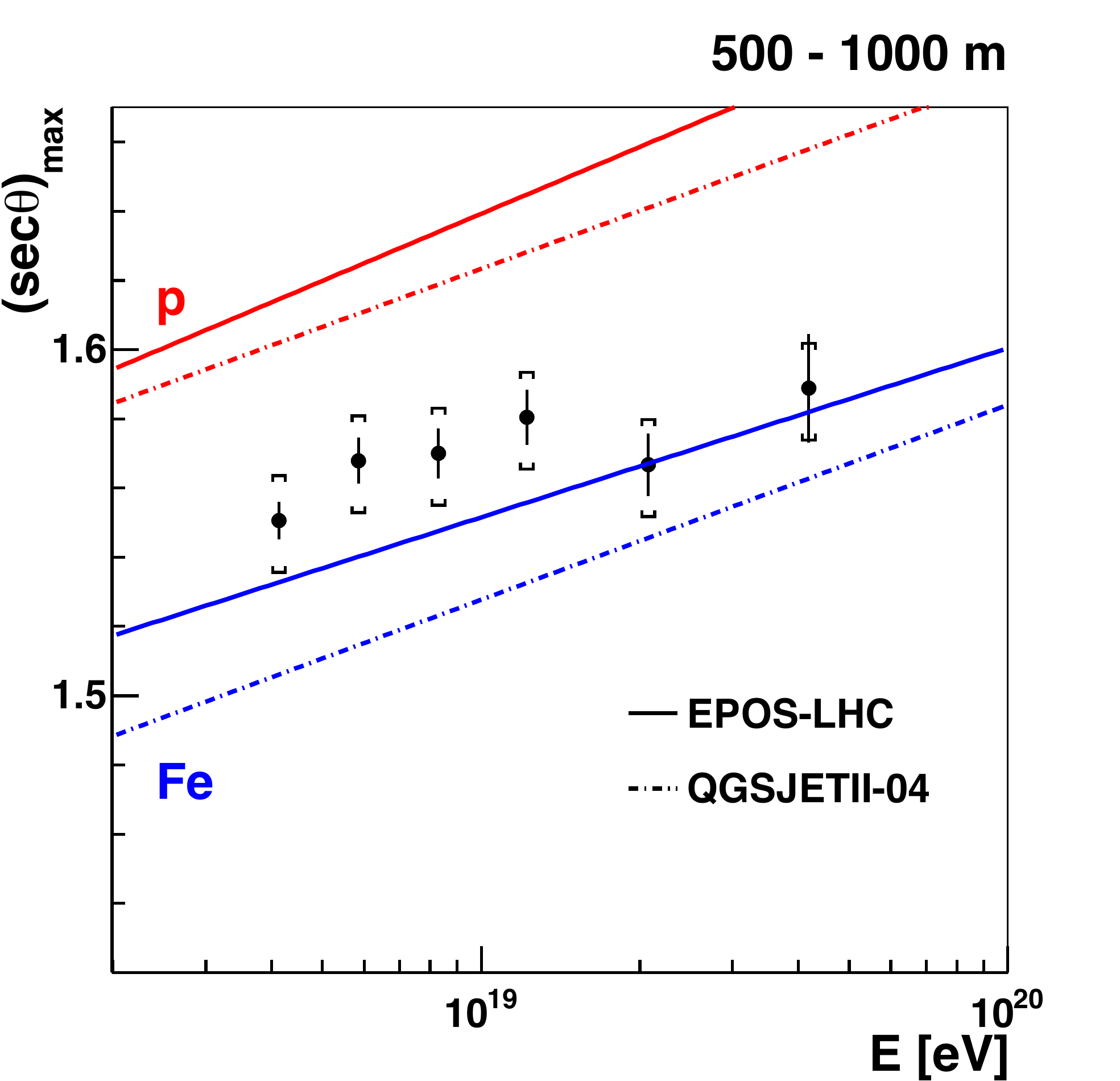}
        \includegraphics[width=0.45\textwidth]{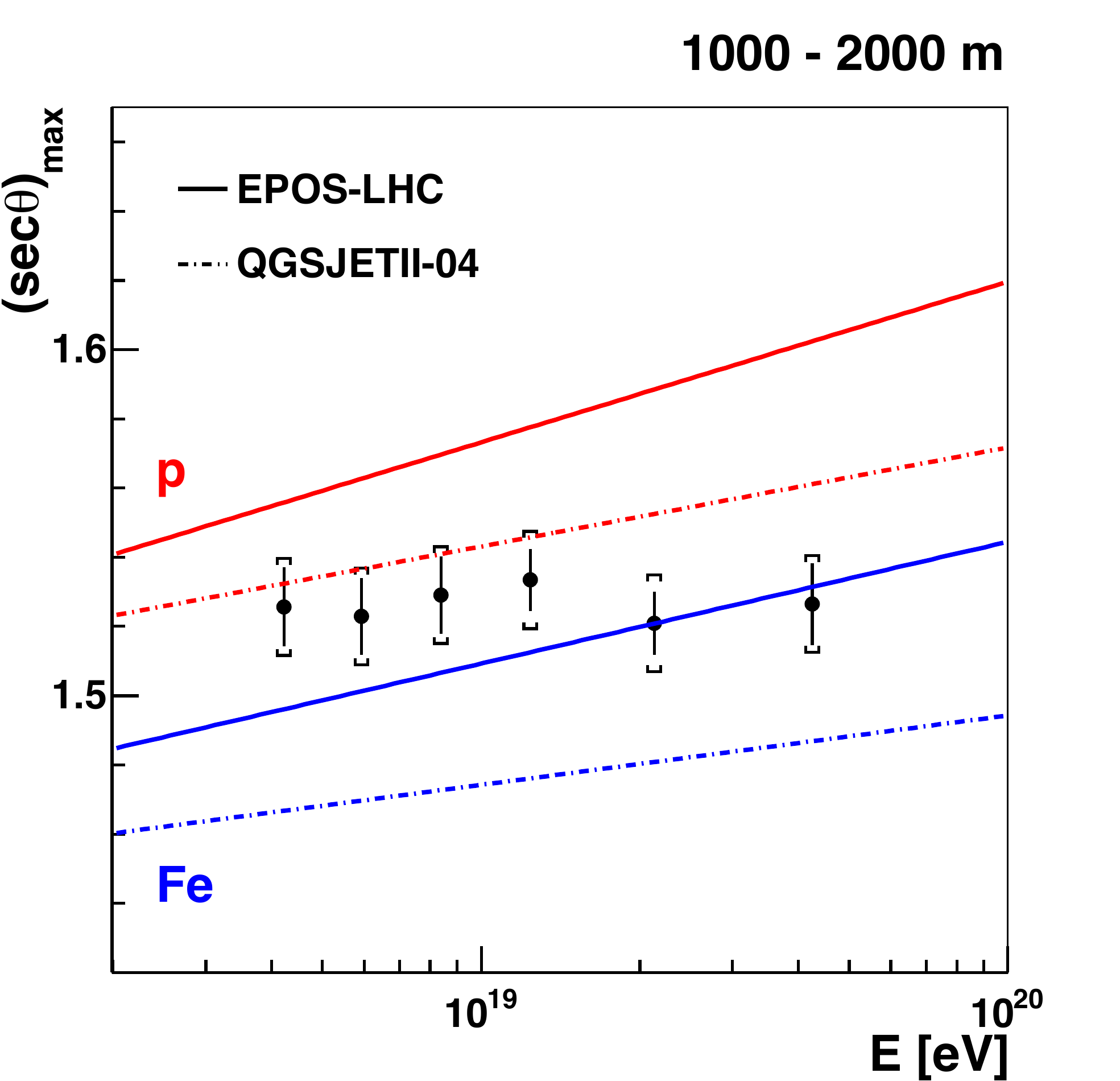}
        \caption{Comparison between $\secMax$, for both data and Monte Carlo predictions in the 500 $-$ 1000 m interval (top) and in the 1000 $-$ 2000 m interval (bottom) using both hadronic models EPOS-LHC (solid lines) and QGSJETII-04 (dashed lines), for both
primaries, proton (red) and iron (blue).}\label{fig:sec_max_MC}
  \end{center}
\end{figure}

The comparison of the energy dependence of the measured $\secMax$ with predictions for proton and iron primaries, and for both hadronic models, is shown in Fig.~\ref{fig:sec_max_MC}.
The systematic uncertainty on the measured $\secMax$ is 16$\%$ (500 $-$ 1000 m) and 21$\%$ (1000 $-$ 2000 m) of the predicted separation between proton-iron $\secMax$ for both models. From this figure it is evident that the Auger data are bracketed by proton and iron in both models, independent of the core distance interval studied. However, the dependence of $\secMax$ on energy is such that it is difficult to draw strong conclusions as rather different predictions come from the two models, particularly in the larger distance interval. However, in both cases there is an indication that the mean mass increases slowly with energy in line with other Auger studies \cite{Aab:2014kda,MPD:2014dua}. 

It is also evident from these plots that the mass predictions depend strongly on the hadronic model adopted. To study these discrepancies further, we have transformed the measurements of $\secMax$ (and their corresponding uncertainties) into mass units. 

For each interaction model, the value of $\MeanlnA$ derived from data has been computed using the following relationships:

\begin{equation}
  \lnA = \frac{\secMaxp-(\sec\theta)_{\rm max;data}}{\secMaxp-\secMaxFe}\cdot \ln 56 \label{eq:LnA}
\end{equation}

\begin{equation}
  \Delta \lnA = -\frac{\Delta(\sec\theta)_{\rm max;data}}{\secMaxp - \secMaxFe}\cdot \ln 56 \label{eq:DeltaLnA}
\end{equation}

\begin{figure}[t!]
  \begin{center}
        \includegraphics[width=0.49\textwidth]{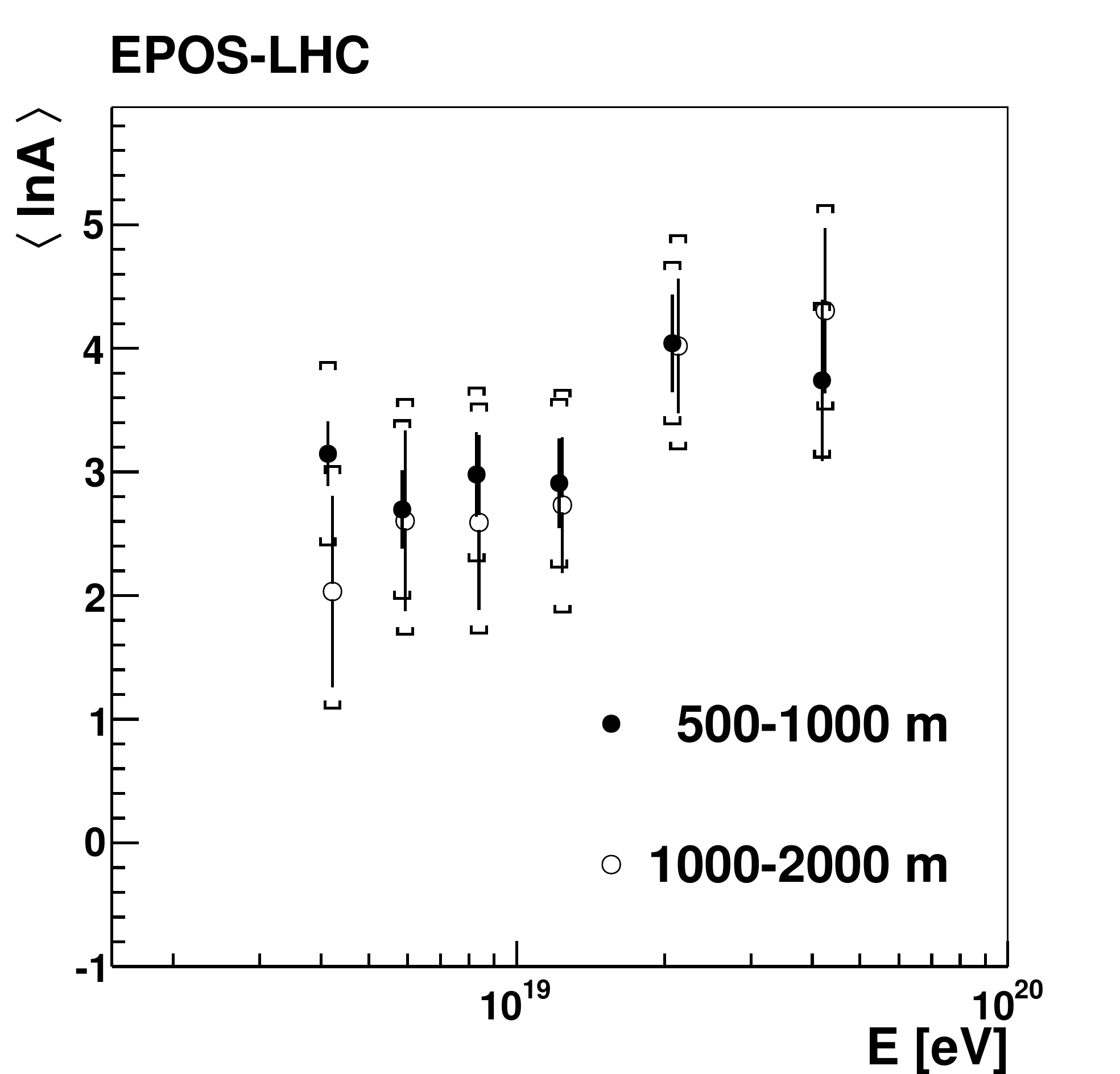}
        \includegraphics[width=0.49\textwidth]{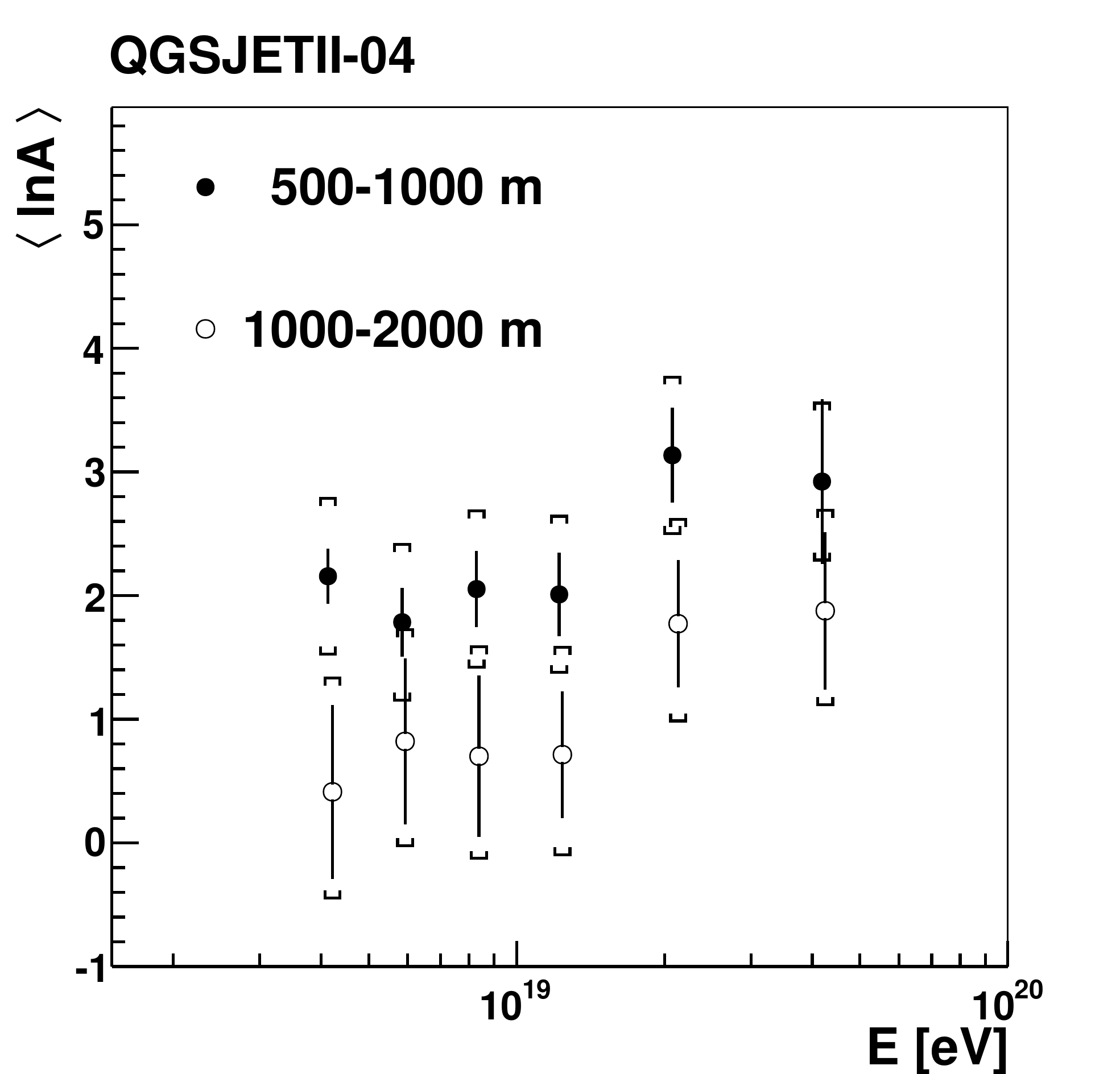}
        \caption{Comparison of $\MeanlnA$ as a function of energy for both core distance intervals predicted by EPOS-LHC (top panel) and QGSJETII-04 (bottom panel).}\label{fig:comparisonlna}
  \end{center}
\end{figure}

The result of this transformation is shown in Fig.~\ref{fig:comparisonlna}. While for the EPOS-LHC model the mean mass is independent of the radial interval used in the analysis, as expected, this is much less evident for the QGSJETII-04 model.
These results imply that the study of $\secMax$ can also be used to probe the validity of hadronic interaction models.

\section{\label{sec:Conclusions}Comparison with previous measurements and conclusions}

The azimuthal dependence of the $\rt$ values obtained from about $2 \times 10^5$ FADC traces registered by the SD detector of the Pierre Auger Observatory has been used to obtain a mass-sensitive parameter, $\secMax$. The evolution of this parameter as a function of energy, above $3 \times 10^{18}$ eV, has been studied in two ranges of core distance interval. The comparison with predictions from the most up-to-date hadronic models, EPOS-LHC and QGSJETII-04, although hinting at a transition from lighter to heavier composition as the energy increases, does not allow us to draw strong conclusions on its absolute value. This is because the predictions are at variance not only with the two models, but even with the two distance ranges. In particular, the comparison between data and predictions from QGSJETII-04 suggests unphysical conclusions, with the mass seemingly dependent upon the distance of the stations from the core. This is a clear indication that further deficiencies in the modelling of showers must be resolved before $\secMax$ can be used to make inferences about mass composition. It also shows that the reach of the $\secMax$ observable extends to providing a test of hadronic interactions models.

\begin{figure*}[t!]
        \includegraphics[width=0.49\textwidth]{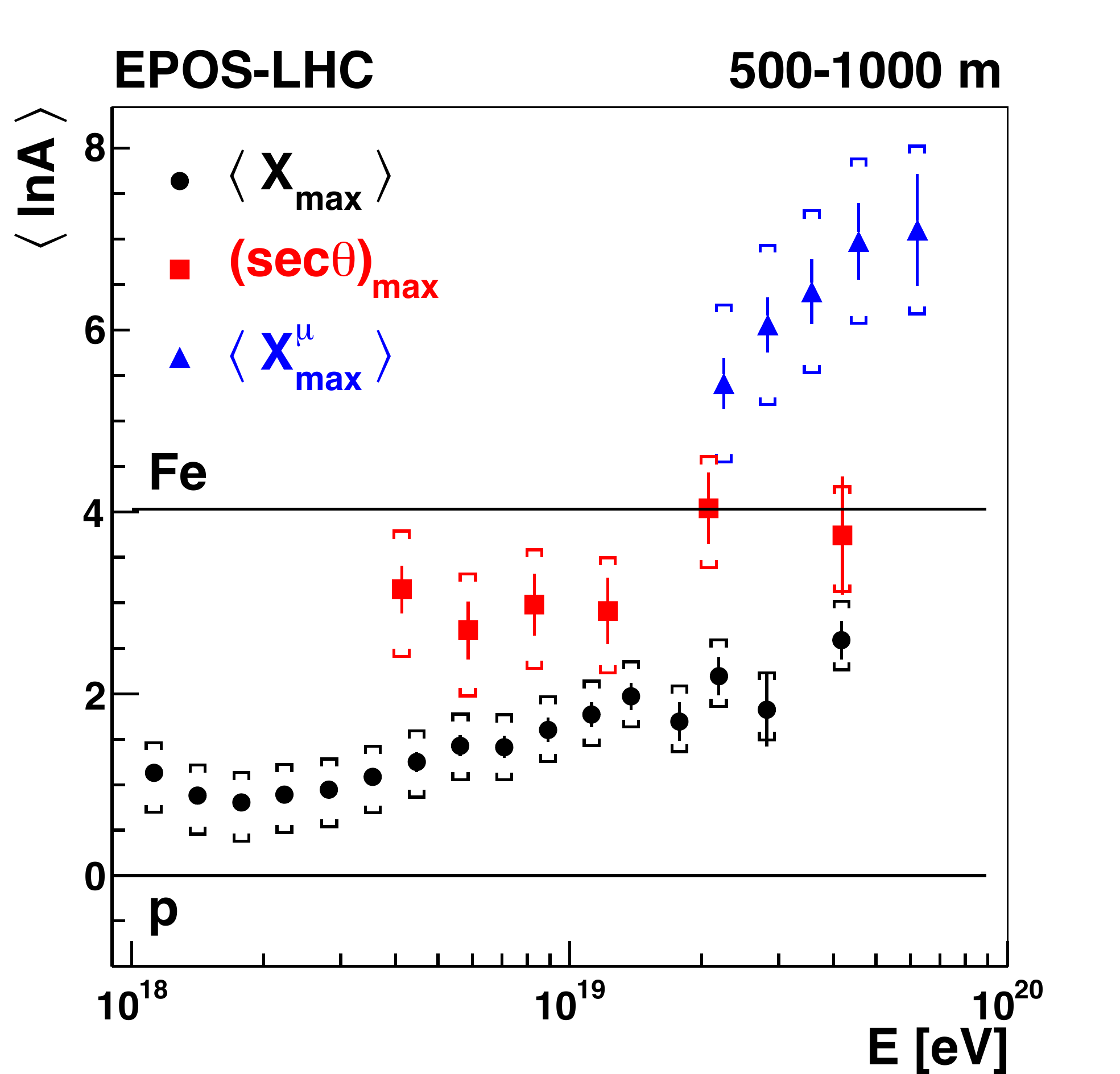}
        \includegraphics[width=0.49\textwidth]{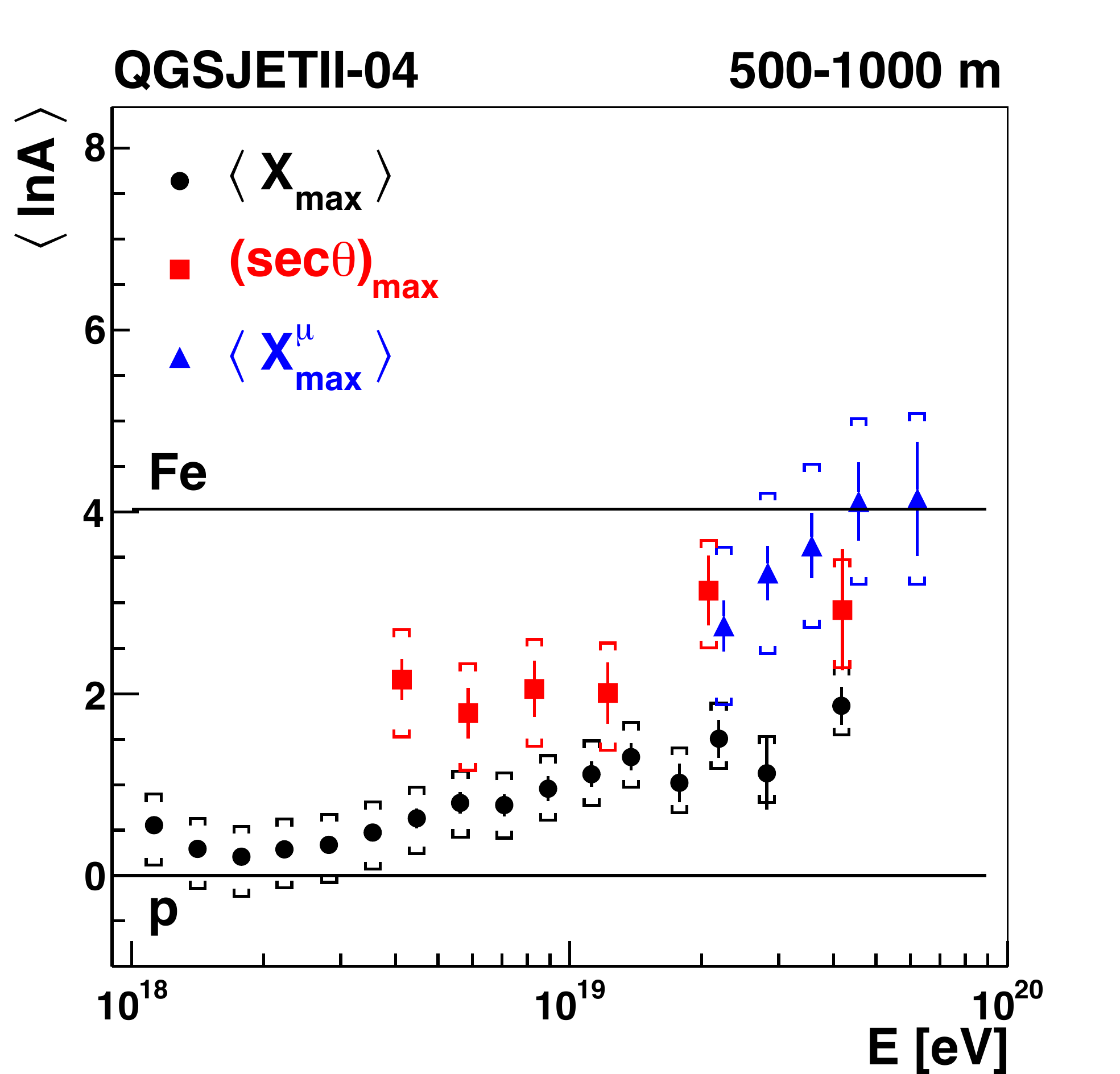}
        \includegraphics[width=0.49\textwidth]{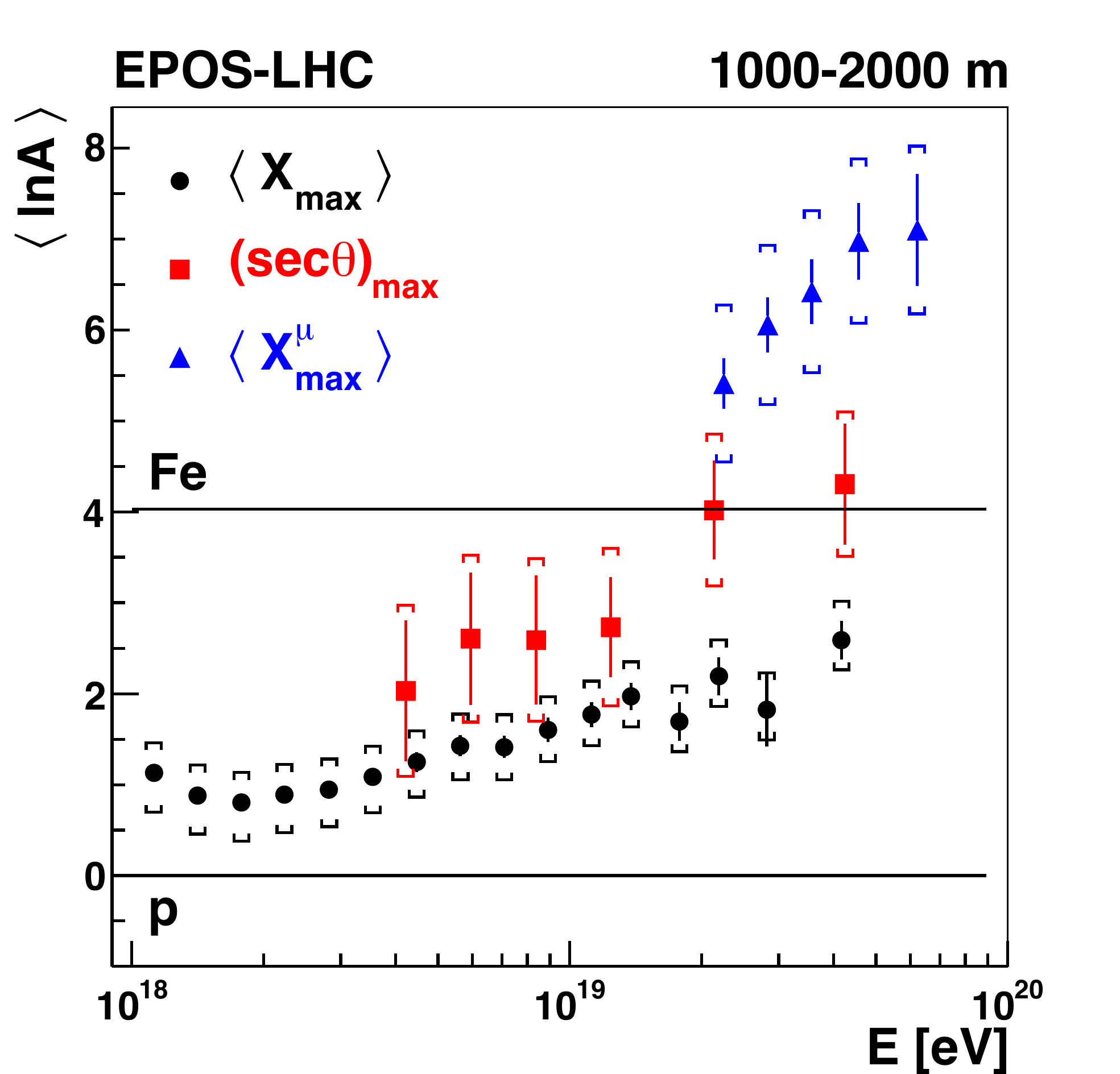}
        \includegraphics[width=0.49\textwidth]{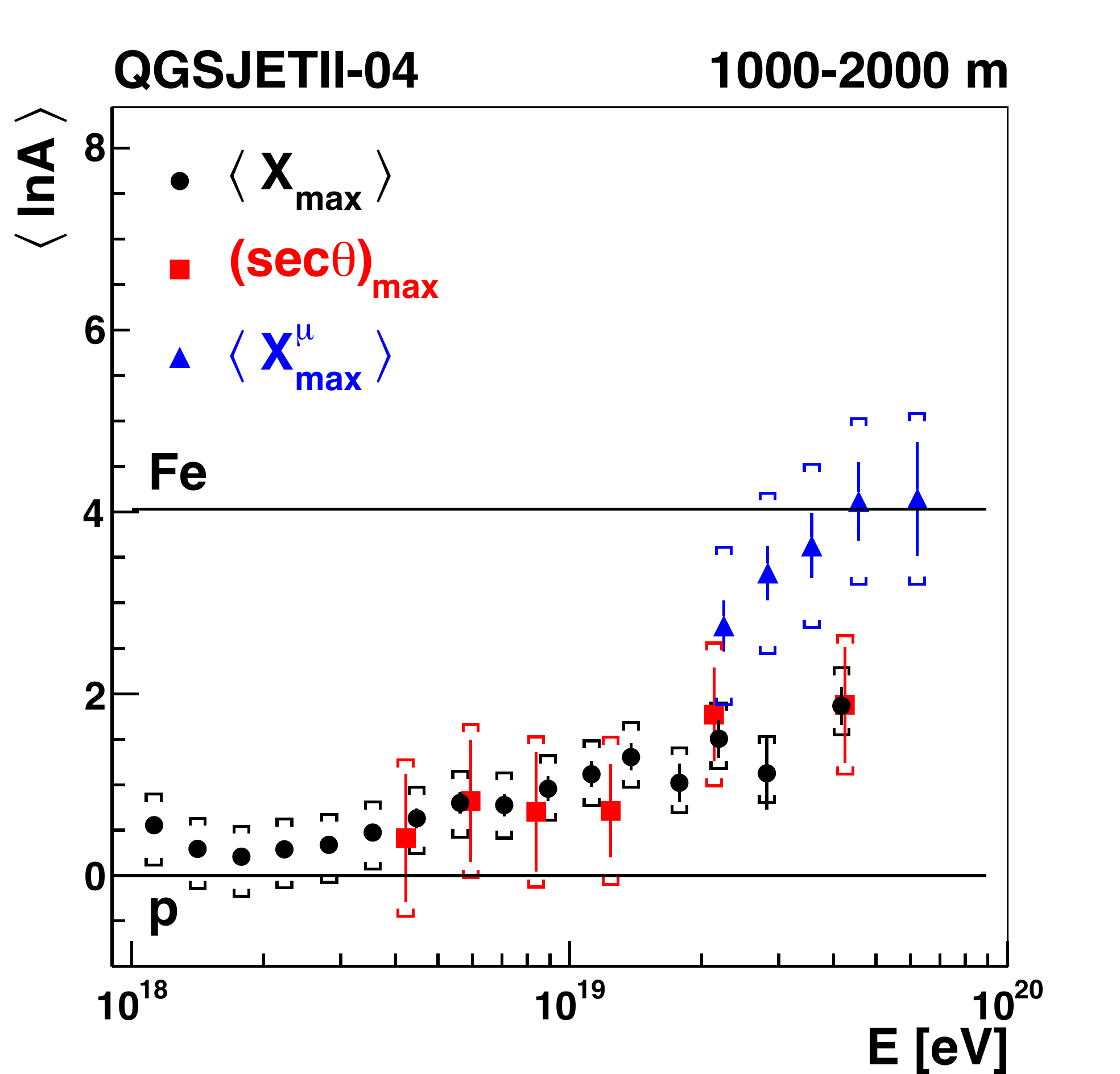}
        \caption{$\MeanlnA$ as a function of energy as predicted by EPOS-LHC and QGSJETII-04. Results from the asymmetry analysis in both $r$ intervals are shown and compared with those from the elongation curve \cite{Aab:2014kda} and the MPD 
method \cite{MPD:2014dua}.}\label{fig:meanlna}
\end{figure*}

We conclude by making a comparison in Fig.~\ref{fig:meanlna} of mass values (in terms of $\MeanlnA$) obtained from the measurements of
$\secMax$ for the two distance ranges to previous mass estimates from the Pierre Auger Observatory \cite{Aab:2014kda,MPD:2014dua}.
The three mass measurements have different systematic uncertainties and are sensitive to very different types of hadronic interactions since the importance of the muonic shower
component is different within each of them. In the direct determination of $\Xmax$ \cite{Aab:2014kda}, the dominant shower component is the electromagnetic one and the proportion 
of muons in the shower is of minor importance. As a consequence in that case the dominant contribution comes from the very first high energy hadronic interactions 
\cite{Ulrich:2015ipa}. By contrast, the muon production-depth \cite{MPD:2014dua} is dominated by the muon component which is the result of a long cascade of lower energy hadronic 
interactions (mostly pion-nucleus interactions) \cite{Ostapchenko:2016bir}. The asymmetry in the risetime is associated with a complex interplay between these two components. As 
these three measurements lead to discordant estimates of $\MeanlnA$, it is impossible to conclude which of the two models considered here best describes the totality of the data. 
While the EPOS model yields results that are consistent at different distances (Fig.~\ref{fig:comparisonlna}) for instance, the mass values predicted from the muon production-depth 
(Fig.~\ref{fig:meanlna}) would imply that trans-uranic elements are dominant above 20~EeV. The $\langle X_\text{max}^{\mu} \rangle$ result, and a related analysis of muons in very 
inclined showers made at the Auger Observatory \cite{Aab:2014pza}, suggest that the muon component of showers is incorrectly modelled. In particular, the measured pion-carbon 
cross-section for the production of a forward $\rho^0$ meson, which decays to two charge pions, instead of $\pi^0$ as leading particle exceeds what has been included in the models 
\cite{Herve:2015lra} and work is underway to evaluate the importance of this effect on muon production and MPD. Moreover the lack of measurements of the production of forward 
baryons in pion-nucleus interactions, which also has a large effect on muon production \cite{Pierog:2006qv} and on $\langle X_\text{max}^{\mu} \rangle$ \cite{Ostapchenko:2016bir}, 
leads to large uncertainties in model predictions. Additionally one must not overlook the possibility that a new phenomenon, such as described in 
\cite{Allen:2013hfa,Farrar:2013sfa}, could become important at the energies studied here which explore the centre-of-mass region well above that studied directly at the LHC. 
Discriminating between such possibilities is a target of the AugerPrime project \cite{Aab:2015bza} which will have the ability to separate the muon and electromagnetic signals.



\section*{Acknowledgments}

\begin{sloppypar}
The successful installation, commissioning, and operation of the Pierre Auger
Observatory would not have been possible without the strong commitment and
effort from the technical and administrative staff in Malarg\"ue. We are very
grateful to the following agencies and organizations for financial support:
\end{sloppypar}

\begin{sloppypar}
Comisi\'on Nacional de Energ\'\i{}a At\'omica, Agencia Nacional de Promoci\'on Cient\'\i{}fica y Tecnol\'ogica (ANPCyT), Consejo Nacional de Investigaciones Cient\'\i{}ficas y T\'ecnicas (CONICET), Gobierno de la Provincia de Mendoza, Municipalidad de Malarg\"ue, NDM Holdings and Valle Las Le\~nas, in gratitude for their continuing cooperation over land access, Argentina;
the Australian Research Council;
Conselho Nacional de Desenvolvimento Cient\'\i{}fico e Tecnol\'ogico (CNPq), Financiadora de Estudos e Projetos (FINEP), Funda\c{c}\~ao de Amparo \`a Pesquisa do Estado de Rio de Janeiro (FAPERJ), S\~ao Paulo Research Foundation (FAPESP) Grants No.\ 2010/07359-6 and No.\ 1999/05404-3, Minist\'erio de Ci\^encia e Tecnologia (MCT), Brazil;
Grant No.\ MSMT-CR LG13007, No.\ 7AMB14AR005, and the Czech Science Foundation Grant No.\ 14-17501S, Czech Republic;
Centre de Calcul IN2P3/CNRS, Centre National de la Recherche Scientifique (CNRS), Conseil R\'egional Ile-de-France, D\'epartement Physique Nucl\'eaire et Corpusculaire (PNC-IN2P3/CNRS), D\'epartement Sciences de l'Univers (SDU-INSU/CNRS), Institut Lagrange de Paris (ILP) Grant No.\ LABEX ANR-10-LABX-63, within the Investissements d'Avenir Programme Grant No.\ ANR-11-IDEX-0004-02, France;
Bundesministerium f\"ur Bildung und Forschung (BMBF), Deutsche Forschungsgemeinschaft (DFG), Finanzministerium Baden-W\"urttemberg, Helmholtz Alliance for Astroparticle Physics (HAP), Helmholtz-Gemeinschaft Deutscher Forschungszentren (HGF), Ministerium f\"ur Wissenschaft und Forschung, Nordrhein Westfalen, Ministerium f\"ur Wissenschaft, Forschung und Kunst, Baden-W\"urttemberg, Germany;
Istituto Nazionale di Fisica Nucleare (INFN),Istituto Nazionale di Astrofisica (INAF), Ministero dell'Istruzione, dell'Universit\'a e della Ricerca (MIUR), Gran Sasso Center for Astroparticle Physics (CFA), CETEMPS Center of Excellence, Ministero degli Affari Esteri (MAE), Italy;
Consejo Nacional de Ciencia y Tecnolog\'\i{}a (CONACYT), Mexico;
Ministerie van Onderwijs, Cultuur en Wetenschap, Nederlandse Organisatie voor Wetenschappelijk Onderzoek (NWO), Stichting voor Fundamenteel Onderzoek der Materie (FOM), Netherlands;
National Centre for Research and Development, Grants No.\ ERA-NET-ASPERA/01/11 and No.\ ERA-NET-ASPERA/02/11, National Science Centre, Grants No.\ 2013/08/M/ST9/00322, No.\ 2013/08/M/ST9/00728 and No.\ HARMONIA 5 -- 2013/10/M/ST9/00062, Poland;
Portuguese national funds and FEDER funds within Programa Operacional Factores de Competitividade through Funda\c{c}\~ao para a Ci\^encia e a Tecnologia (COMPETE), Portugal;
Romanian Authority for Scientific Research ANCS, CNDI-UEFISCDI partnership projects Grants No.\ 20/2012 and No.\ 194/2012, Grants No.\ 1/ASPERA2/2012 ERA-NET, No.\ PN-II-RU-PD-2011-3-0145-17 and No.\ PN-II-RU-PD-2011-3-0062, the Minister of National Education, Programme Space Technology and Advanced Research (STAR), Grant No.\ 83/2013, Romania;
Slovenian Research Agency, Slovenia;
Comunidad de Madrid, FEDER funds, Ministerio de Educaci\'on y Ciencia, Xunta de Galicia, European Community 7th Framework Program, Grant No.\ FP7-PEOPLE-2012-IEF-328826, Spain;
Science and Technology Facilities Council, United Kingdom;
Department of Energy, Contracts No.\ DE-AC02-07CH11359, No.\ DE-FR02-04ER41300, No.\ DE-FG02-99ER41107 and No.\ DE-SC0011689, National Science Foundation, Grant No.\ 0450696, The Grainger Foundation, USA;
NAFOSTED, Vietnam;
Marie Curie-IRSES/EPLANET, European Particle Physics Latin American Network, European Union 7th Framework Program, Grant No.\ PIRSES-2009-GA-246806;
and UNESCO.
\end{sloppypar}

\bibliographystyle{apsrev4-1}
\bibliography{asymprd}

\end{document}